\documentclass[preprint,12pt,authoryear]{elsarticle}

\usepackage{amssymb}
\usepackage{rpdpreamble_3}
\pdfoutput=1
\usepackage{graphicx}
\usepackage{epstopdf, epsfig}
\usepackage{enumerate}
\usepackage{subcaption}
\usepackage{xcolor}
\usepackage{setspace}

\journal{Computers and Fluids}

\begin{document}

\begin{frontmatter}



\title{Data-Driven Modelling of the Reynolds Stress Tensor using Random Forests with Invariance}


\author[label1,label2]{Mikael L.A. Kaandorp}
\ead{m.l.a.kaandorp@uu.nl}
\author[label1]{Richard P. Dwight}
\address[label1]{Aerodynamics, Faculty of Aerospace, Delft University of Technology, 2629 HS, Delft, The Netherlands}
\address[label2]{Institute for Marine and Atmospheric research Utrecht (IMAU), Utrecht University, 3584 CC, Utrecht, The Netherlands}


\begin{abstract}
A novel machine learning algorithm is presented, serving as a data-driven turbulence modeling tool for Reynolds Averaged Navier-Stokes (RANS) simulations. This machine learning algorithm, called the Tensor Basis Random Forest (TBRF), is used to predict the Reynolds-stress anisotropy tensor, while guaranteeing Galilean invariance by making use of a tensor basis. By modifying a random forest algorithm to accept such a tensor basis, a robust, easy to implement, and easy to train algorithm is created.  The algorithm is trained on several flow cases using DNS/LES data, and used to predict the Reynolds stress anisotropy tensor for new, unseen flows.  The resulting predictions of turbulence anisotropy are used as a turbulence model within a custom RANS solver.  Stabilization of this solver is necessary, and is achieved by a continuation method and a modified $k$-equation.   Results are compared to the neural network approach of Ling et al. [J. Fluid Mech, 807(2016):155-166, (2016)]. Results show that the TBRF algorithm is able to accurately predict the anisotropy tensor for various flow cases, with realizable predictions close to the DNS/LES reference data. Corresponding mean flows for a square duct flow case and a backward facing step flow case show good agreement with DNS and experimental data-sets.  Overall, these results are seen as a next step towards improved data-driven modelling of turbulence. This creates an opportunity to generate custom turbulence closures for specific classes of flows, limited only by the availability of LES/DNS data.

\end{abstract}




\end{frontmatter}


\section{Introduction}
\label{sec:intro}
The last few years have seen the introduction of supervised machine learning (ML) algorithms as tools to exploit data for the purpose of modeling turbulence.  RANS models are currently, and are expected to remain the norm for simulating turbulent flows in most industrial applications~\citep{Slotnick2014}, because of their computational tractability, but suffer from poor accuracy and predictive power in a variety of important flows  \citep{Craft1996}.  While a variety of nonlinear eddy-viscosity models (NLEVMs) and Reynolds-stress transport (RST) models have been developed using traditional techniques, it is the simplest linear eddy viscosity models such as the $k-\epsilon$ model and $k-\omega$ models that remain by far the most widely used.  This has motivated some to advocate alternative modelling approaches that utilize available reference data more fully, and rely less on physical reasoning \citep{Duraisamy2019}.  Supervised machine-learning methods developed in other fields, are -- in the best case -- able to distill large data-sets into simple functional relationships.  This offers the hope of substantially improving current RANS models, by building closures customized for a particular class of flows based on appropriate reference LES or DNS data-sets \citep{Ling2016,Wang2017}. However there exist significant obstacles to realizing these models in practice.

Firstly, a high sensitivity of the mean-flow to the detailed character of the turbulence has been reported -- even in the case of channel flows \citep{Thompson2016}.  This places stringent accuracy requirements on any data-derived closure model.  Secondly, there exists no unique map from local mean-flow quantities to the needed turbulence statistics, due to non-local and non-equilibrium physics.  Thirdly, any closure model should incorporate basic physical constraints, such as Galilean invariance: readily achievable in analytically derived models, but difficult when employing ML procedures which generate arbitrary functions.  Fourthly, a ML model should produce smooth fields (they must be incorporated into a PDE solver), yet able to capture flows with rapid spatial variations, as in e.g.\ boundary-layers.  Finally, RANS solvers are notoriously unstable and difficult to drive to convergence, even with standard closure models.  Stabilization of a data-derived model is therefore necessary.  These challenges are in addition to the standard ML challenges. Think for example training against large volumes of data with high dimensionality \citep{Chen2014}, and avoiding overfitting, which can be done by using for example cross-validation methods, or methods specific to the algorithm used, e.g. randomly dropping connections in neural networks \citep{Srivastava2014}, or simplifying decision trees by pruning them back \citep{Esposito1997}.

In this paper, a new approach is presented that addresses all these challenges to some extent, resulting in a closure model that significantly outperforms a baseline RANS model for a specified class of flows.  The model is closely related to nonlinear eddy-viscosity models (NLEVMs), of e.g.\ \cite{Pope1975}, in which the normalized Reynolds-stress anisotropy tensor is predicted from the local mean-flow.  We integrate a tensor basis for the anisotropy into a modified random-forest method, resulting in the Tensor Basis Random Forest (TBRF), analogously to the Tensor Basis Neural Network (TBNN) of \cite{Ling2016}.  Galilean invariance can therefore be guaranteed; and in comparison to artificial neural networks, our random forests are easier to implement, easier to train, have fewer hyperparameters, and have natural techniques for avoiding overfitting \citep{Hastie2008}.  We introduce a method for stabilizing the RANS solver, using a combination of a modified $k$-equation, and relaxation against a Boussinesq stress-tensor.  With this solver we can converge mean-flows to a steady state, with our TBRF closure model.

Many types of approach can be identified in the literature for improvements of RANS closure models with data, we only give a selection here.  Broadly speaking there are parametric approaches, which calibrate or slightly modify existing models as in \citep{Cheung2011,Edeling2014} (often with statistical inference); and structural approaches, which attempt to relax fundamental modelling assumptions, especially Boussinesq as in \citep{Ling2016,Wang2010}.  In the latter case, uncertainty quantification has been used to develop predictive models in the absence of data, by incorporating stochastic terms intended to capture the effect of modelling assumptions, see \citep{Tracey2013,Emory2013,Xiao2016}.

With data available, machine-learning has been used to capture potentially complex relationships between mean-flow and modelling terms.  These approaches can largely be divided into those which solve inverse problems to identify local correction terms needed to match data, as in \cite{Duraisamy2015}, \cite{Parish2016}, and \cite{Singh2016a}; and those which apply machine-learning to directly predict model terms based on local mean-flow only, see \cite{Ling2015}.  The machine-learning methods used are various, \cite{Ling2016} use neural networks; \cite{Wang2017,Ling2016a} use random forests; and \cite{Weatheritt2016} uses gene-expression programming to obtain a concise form of their model.  Despite the popularity of random-forests, existing authors have not incorporated frame-invariance, or solver stabilization that we introduce here.  These developments are critical for the robustness and practicality of the method.

This paper is structured as follows. Section \ref{sec:methodology} will discuss the methodology for the TBRF framework, which includes underlying theory, the implementation of the algorithm itself, and the data-sets used in the framework. Section \ref{sec:results} will discuss the results from the framework, which include predictions for the anisotropy tensor, flow fields obtained after propagating these predictions into the flow field, and some proposals for future work with regards to quantifying uncertainty of the model output. Section \ref{sec:conclusions} will present the conclusions. 

\section{Methods}
\label{sec:methodology}
When performing Reynolds-averaging of the incompressible Navier-Stokes equations, the complete effect of the unresolved turbulent fluctuations on the mean-flow is contained in a single term,
$\nabla\cdot\taub$ (where $[\taub]_{ij} := \overline{u_i' u_j'}$ is the Reynolds-stress tensor), which modifies the Navier-Stokes momentum equation \citep{Reynolds1895}.  To obtain a turbulence closure in this setting, it is sufficient to specify $\taub$ in terms of mean-flow quantities $\bar u$ etc. The methodological goal of this work is to use plentiful DNS/LES data to estimate a nonlinear eddy-viscosity model (NLEVM) of the general form
\[
\taub \simeq \hf(\nabla \bar\ub, \dots),
\]
where $\hf(\cdot)$ is a function of mean-flow quantities only.  This problem can be cast as a standard supervised machine learning task \citep{Murphy2012}.  However we demand additionally that $\hf$ is invariant to the choice of Galilean coordinate frame, and that the training process is computationally cheap and robust.  The first is achieved with an integrity basis (Section \ref{subsec:invariance}), and the second by using a modified random forest  (Section \ref{subsec:TBRF}).  The model for $\taub$ thus obtained is used within a RANS solver to predict the mean-flow; this procedure requires solver stabilization (described in Section \ref{subsec:NumSetup}).  Finally, training and validation cases with DNS/LES data are listed in Section \ref{subsec:FlowCases}.


\subsection{Outline of framework for data-driven turbulence modeling}
\label{subsec:flowDiagram}
\begin{figure}
\centering
    \includegraphics[trim={0.5cm 5cm 4cm 5cm},clip,width=1\textwidth]{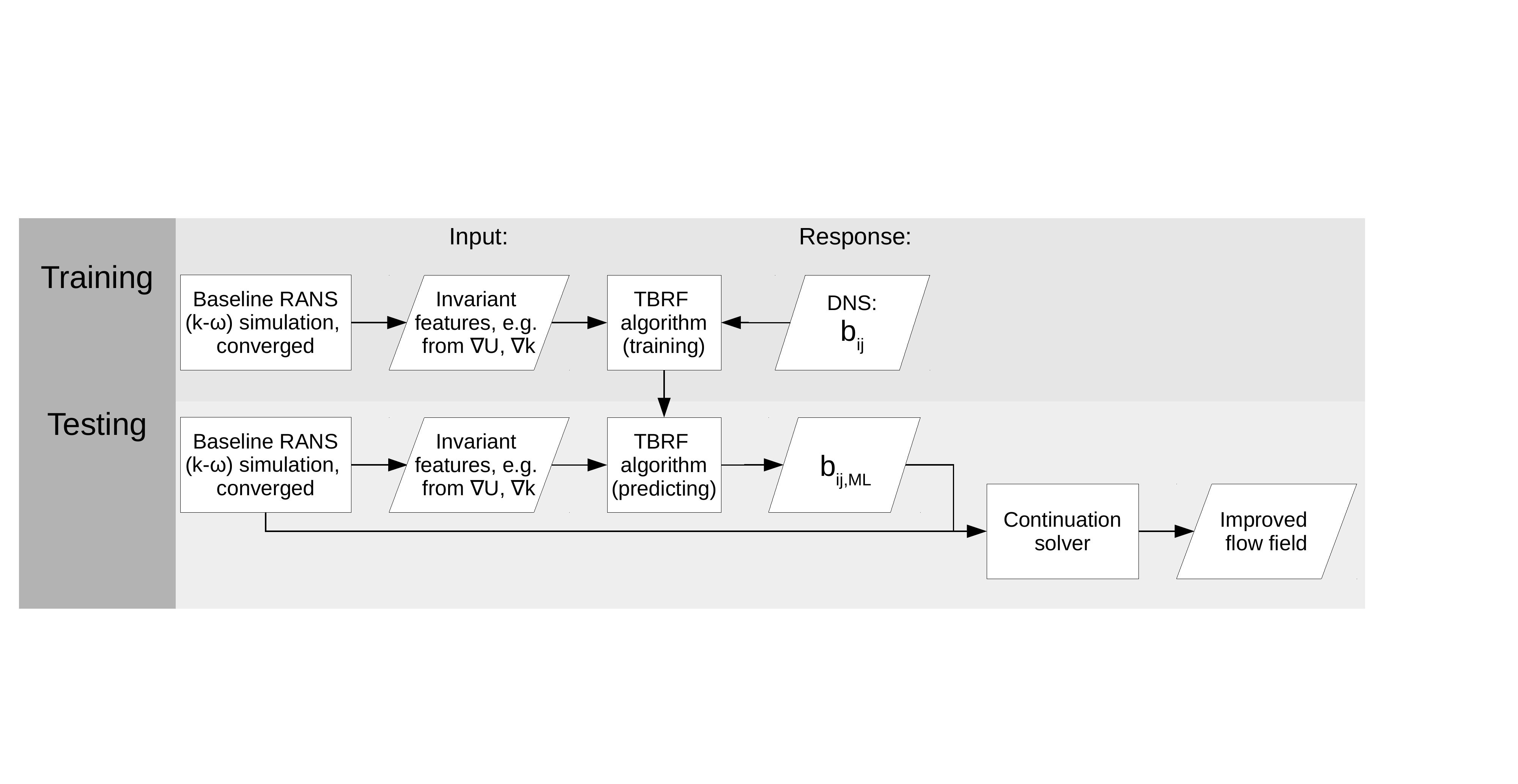}  
    \caption{Flow chart of the machine learning framework.}
    \label{fig:flowDiagram}
\end{figure}

Our framework consists of a training phase and a prediction phase, see Figure~\ref{fig:flowDiagram}.  In the training phase, high-fidelity DNS/LES data is collected for a number of test-cases.  The data should ideally contain full first- and second-order single-point statistics highly resolved in the spatial domain, from which the normalized anisotropy tensor, $\bfff$, is computed, see Section \ref{subsec:ReStressDecomp}.  These same test-cases are simulated with RANS using the $k-\omega$ closure model, and input features are derived from the mean-flow solution at every node of the spatial mesh.  The machine learning algorithm is then trained to approximate the DNS ground-truth anisotropy tensor, from the RANS inputs over the full solution fields of all training-cases simultaneously.

As we use a greedy algorithm to train the decision-trees, the training cost for a data-set of size $N$ is $\mathcal{O}(N \log^2 N)$, so there is no practical restriction on the number of cases which can be used in training (indeed a much more severe limitation has proven to be the availability of high-quality DNS/LES data).  However, we do not expect the map from the mean-flow to unresolved turbulence statistics to be unique, even for a very limited class of flows.  So, in order to capture this non-uniqueness and to prevent overfitting, multiple decision-trees are trained across random subsets of the data.

In the prediction phase, for a previously unseen flow case, the anisotropy tensor is estimated using the mean of the random forest predictions, with input from a baseline RANS mean-field.  An updated mean-field is obtained by solving a modified RANS system with the momentum equation supplemented by the predicted anisotropy.

We want to stress that this is a corrective approach, with a single ML prediction providing an updated solution, similar to that practiced in \cite{Ling2016a,Wu2016a}.  In other words, it is not a replacement for a 'standard' turbulence model, where the Reynolds stress is assumed to be a specific function of the mean flow properties, which is then iteratively solved until convergence is achieved. Such an iterative approach would in theory also be conceivable, in which ML predictions of $\bfff$ and modified RANS form a closed loop.  In this case the ML should be trained on the DNS mean-field, not the RANS field.  However such ambitious approaches are currently untested, it is unclear under what conditions the coupled system will converge, and whether the converged solution will resemble ground-truth.  This is however an important topic for further research. 

Adding to this point, it should be stressed that the approach here is only tested for steady RANS flow cases. It is untested for unsteady flows such as present in low-pressure turbines, see e.g. \cite{Akolekar2018}. 

The ground truth for $\bfff$ coming from high-quality DNS/LES data will still contain some error due to e.g. dicretization errors, errors due to a finite averaging time, and possibly model errors. The errors in the ground truth are assumed to be small here (compared to the model errors of the machine learning algorithms), and therefore not taken in account. Supervised machine learning methods exist which theoretically could be able to capture uncertainties in the ground truth (see e.g. Gaussian process regression \citep{Rasmussen2006})

\subsection{Reynolds stress and realizability constraints}
\label{subsec:ReStressDecomp}
First we briefly review some basic properties of the Reynolds stresses, relevant for constructing a meaningful ML model.
Firstly $\taub$ can be intrinsically divided into isotropic and anisotropic (deviatoric) parts:
\begin{equation}
\taub = \frac{2}{3}k\If + \afff,
\label{eq:ReStress}
\end{equation}
where $\afff$ is the Reynolds stress anisotropy tensor, $k := \frac{1}{2}\trace(\taub)$ is the turbulent kinetic energy, and $\If$ is the identity. It is the anisotropic part of the Reynolds stresses which is important: only this part is effective in transporting momentum, while the isotropic stresses can be simply absorbed into a modified pressure term \citep{Pope2000}. The {\it non-dimensional} Reynolds stress anisotropy tensor, $\bfff$, is defined as:
\begin{equation}
\bfff := \frac{\taub}{2k} - \frac{1}{3} \If,
\label{eq:Anisotropy}
\end{equation}
and this is the quantity which we attempt to predict with machine learning.  In the remainder of this paper ``anisotropy tensor'' will refer to $\bfff$.  To model $\bfff$, eddy-viscosity closure models typically make the intrinsic assumption that $\bfff$ is function of local mean-flow quantities only.  Linear eddy-viscosity models such as $k-\epsilon$ and $k-\omega$ \citep{Launder1974, Wilcox2008}, go on to make the specific, Boussinesq assumption that $\bfff \simeq \frac{1}{2}\nu_t (\nabla \bar \ub + \nabla \bar \ub^T) = \nu_t \hat{\Sf}$ for some scalar {\it eddy viscosity} $\nu_t(\xb)$, which remains to be specified.  Both the intrinsic and specific assumptions introduce modelling error.  We aim to estimate and reduce the error in the latter with LES databases and machine-learning.

\begin{figure}
\centering
    \begin{subfigure}[ht]{0.32\textwidth}
        \includegraphics[width=\textwidth]{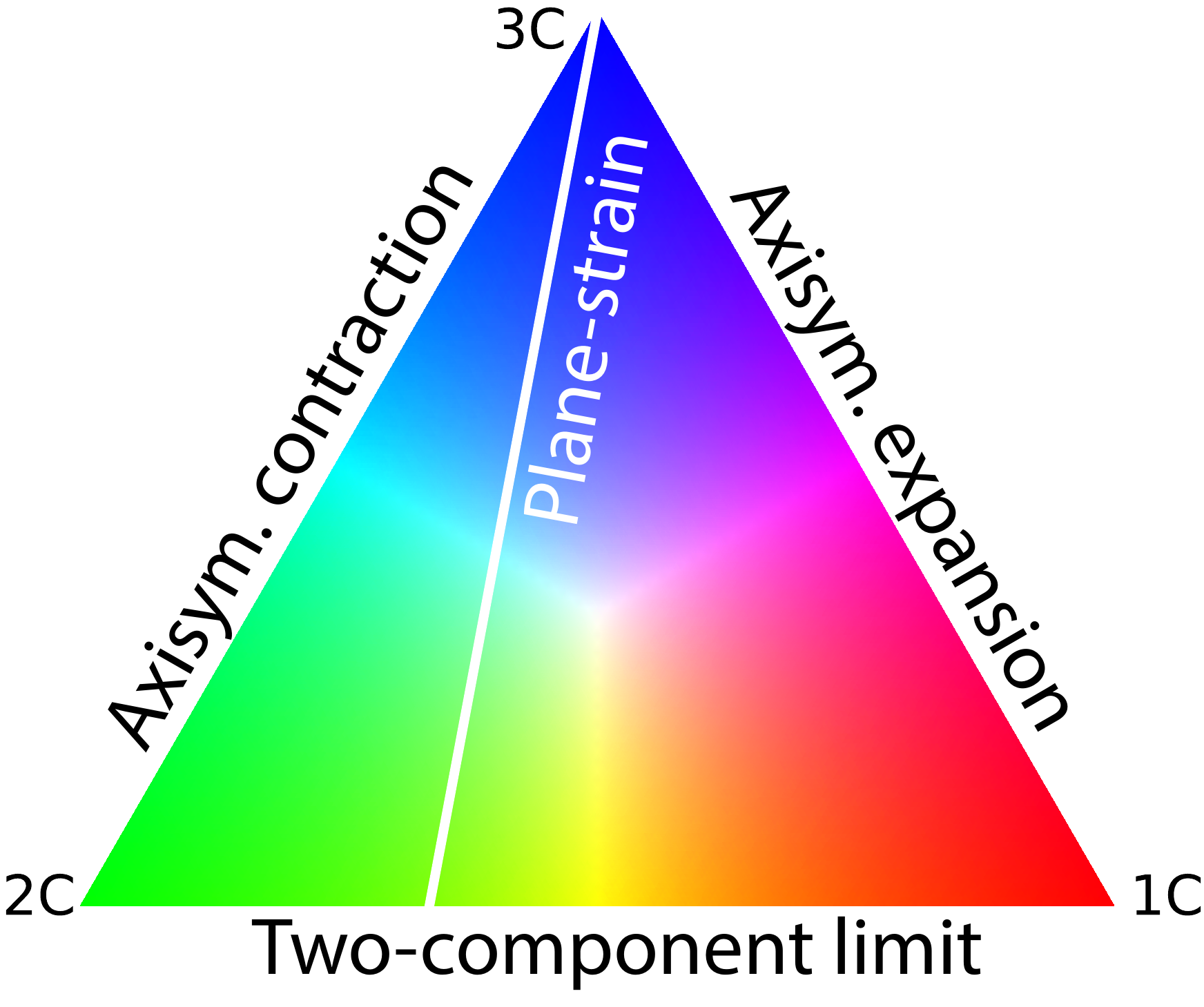}
        \caption{Barycentric map}
        \label{fig:Bary1}
    \end{subfigure}
    \begin{subfigure}[ht]{0.33\textwidth}
        \includegraphics[trim={2cm 0cm 2cm 1.5cm},clip,width=\textwidth]{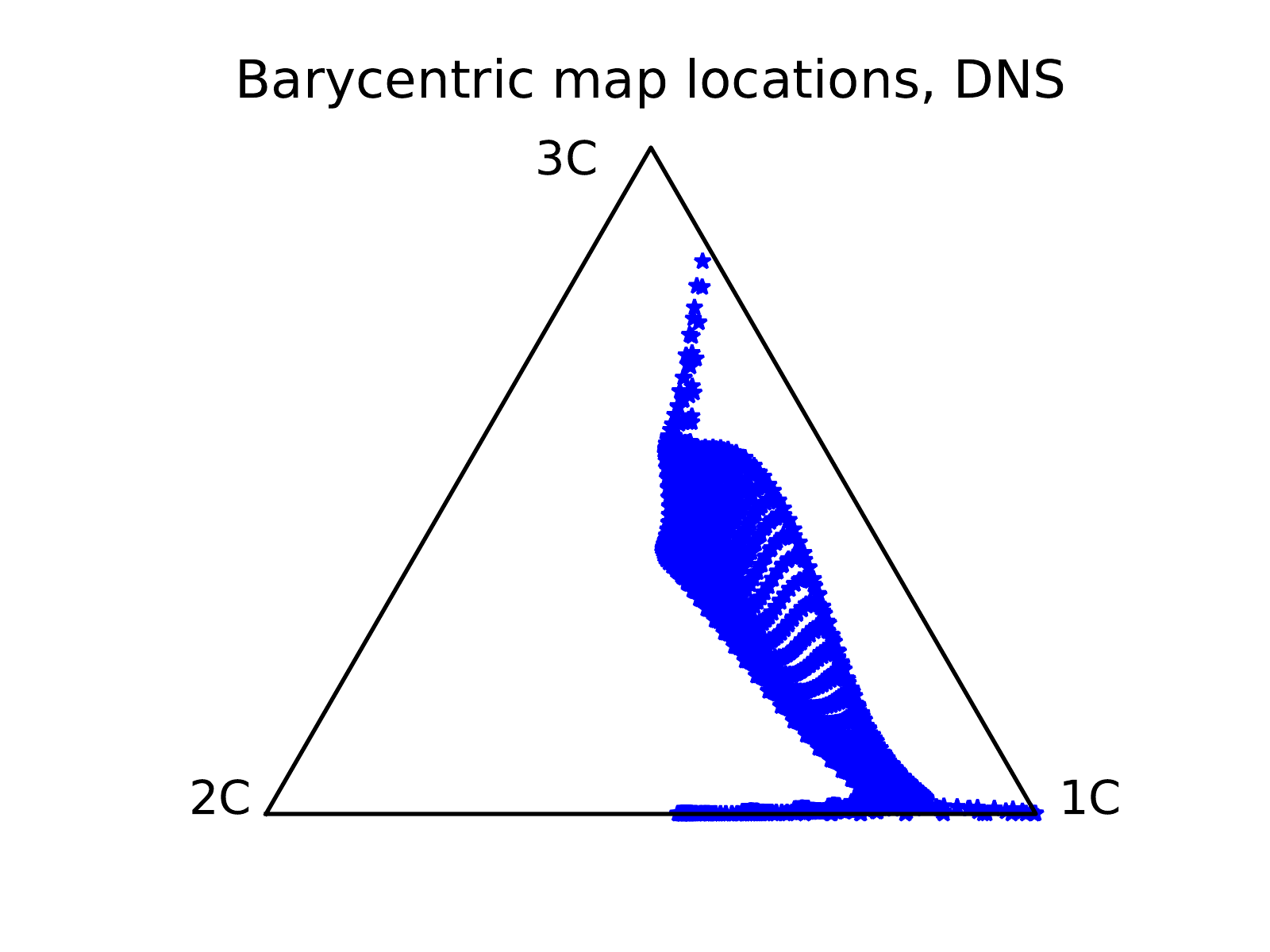}
        \caption{DNS \cite{Pinelli2010}}
        \label{fig:Bary2}
    \end{subfigure} 
    \begin{subfigure}[ht]{0.33\textwidth}
        \includegraphics[trim={2cm 0cm 2cm 1.5cm},clip,width=\textwidth]{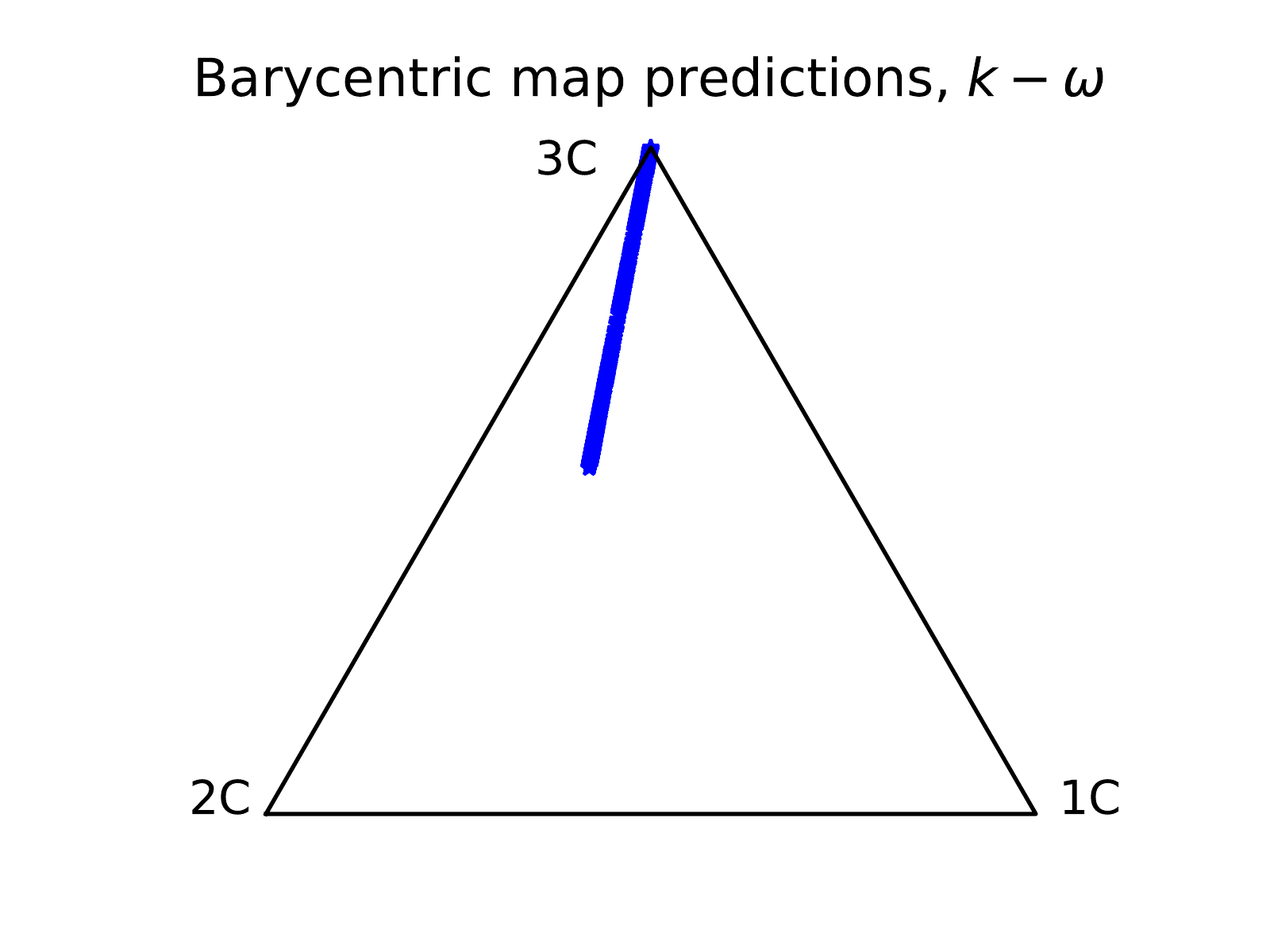}
        \caption{RANS $k-\omega$}
        \label{fig:Bary3}
    \end{subfigure} 
    \caption{Barycentric map (transformation of Lumley triangle), with for a turbulent square duct at $\mathrm{Re} = 3,500$.}
    \label{fig:Bary} 
\end{figure}

The properties of $\taub$ and $\bfff$ lead to physical constraints on models and means of visualization. A matrix $\Af$ is positive semi-definite if (and only if)
$\boldsymbol{x}^T \Af \, \boldsymbol{x} \geq 0,\:\forall \boldsymbol{x}\in\Rbb^N$.
Since the outer product of any vector $\boldsymbol{u'}$ with itself ($\boldsymbol{u'} \otimes \boldsymbol{u'}$) is positive semi-definite; and since the Reynolds stress is an arithmetic average of such tensors, it is also positive semi-definite.  As trivial consequences, all eigenvalues of $\taub$ are real and positive, and 
\begin{equation}
\tau_{\alpha\alpha} \geq 0 \:\:\: \forall \alpha\in\{1,2,3\}, \quad \mathrm{det}(\taub) \geq 0,\quad
\tau_{\alpha\beta}^2 \leq \tau_{\alpha\alpha} \tau_{\beta\beta}\:\:\:\forall \alpha \neq\beta.
\label{eq:posdefcons}
\end{equation}
These properties of $\taub$ have implications for $\bfff$.  Let the eigenvalues of $\taub$ be $\phi_i$, and those of $\bfff$ be $\lambda_i$, then
\begin{equation}
\lambda_i = \frac{\phi_i}{2k} - \frac{1}{3},
\label{eq:eigValsRtoB}
\end{equation}
and both the eigenvalues and diagonal components of $\bfff$ are in the interval $[-\frac{1}{3},\frac{2}{3}]$. Furthermore using the Cauchy-Schwarz inequality in \eqref{eq:posdefcons} the off-diagonal components of $\bfff$ are in $[-\frac{1}{2},\frac{1}{2}]$.  Since $\trace(\bfff)=0$ only two independent invariants of the anisotropy tensor exist, e.g.: $\mathrm{II}:=[\bfff]_{ij}[\bfff]_{ji}$ and $\mathrm{III}:=[\bfff]_{ij}[\bfff]_{in}[\bfff]_{jn}$.  Therefore in the II-III plane all realizable states of turbulence anisotropy can be plotted, which are further restricted to a triangular domain corresponding to the constraints on $\bfff$ just mentioned. This leads to the well-known ``Lumley triangle'' of \cite{Lumley1977,Lumley1979} which captures the anisotropic state of turbulence.  The lesser known barycentric map was introduced in \cite{Banerjee2007}, and is a transformation of the Lumley triangle into barycentric coordinates, and will be used for the purposes of visualization and comparison in this paper, see Figure~\ref{fig:Bary}.

These triangles highlight three limiting states of turbulence anisotropy: 1-component turbulence (restricted to a line, one eigenvalue of $\bfff$ is non-zero), 2-component turbulence (restricted to a plane, two eigenvalues of $\bfff$ are non-zero), and 3-component turbulence (isotropic turbulence, three eigenvalues are non-zero).  Figure~\ref{fig:Bary} shows these, along with invariants for a square-duct flow simulated with DNS from \cite{Pinelli2010}, and a $k-\omega$ RANS simulation.  For any 2d linear eddy-viscosity RANS simulation, the predicted anisotropy invariants will lie entirely along the line indicated as ``plane strain''.  This illustrates the inability of linear eddy-viscosity models to adequately represent anisotropy.

\begin{figure}
\centering
    \begin{subfigure}[ht]{0.24\textwidth}
        \includegraphics[width=\textwidth]{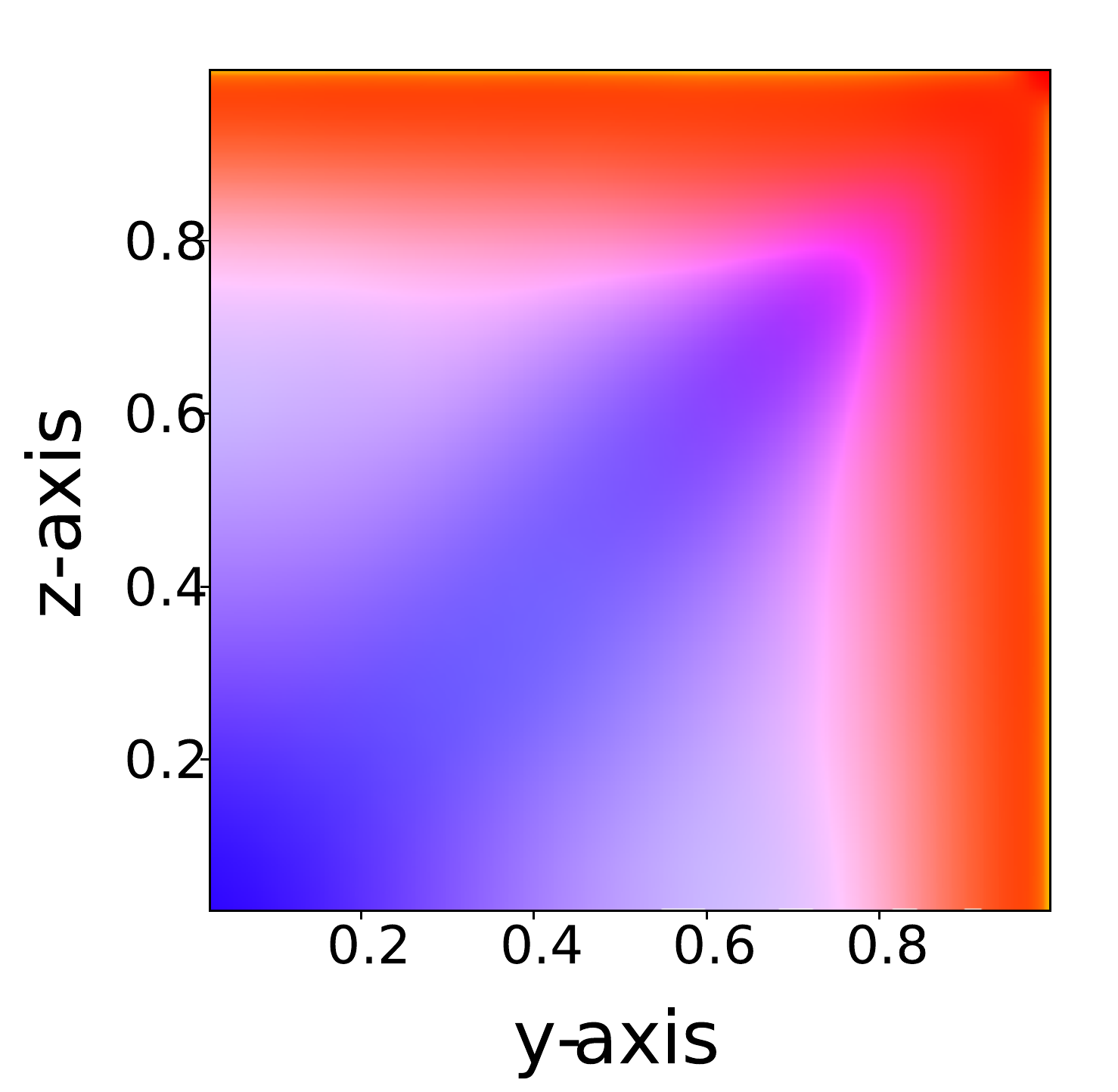}
        \caption{DNS \cite{Pinelli2010}}
        \label{fig:InvariantMap1}
    \end{subfigure}
        \begin{subfigure}[ht]{0.24\textwidth}
        \includegraphics[width=\textwidth]{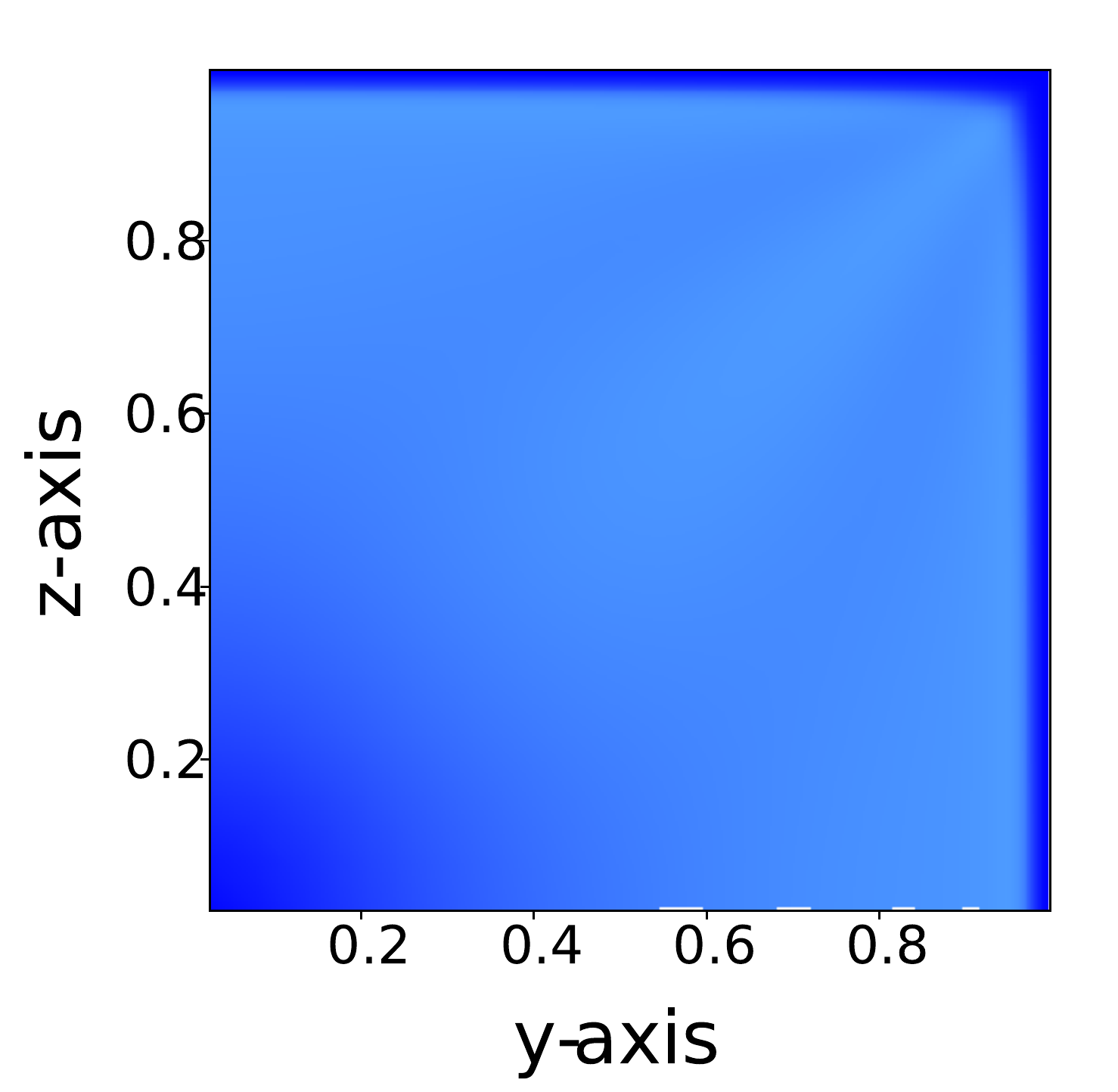}
        \caption{RANS $k-\omega$}
        \label{fig:InvariantMap2}
    \end{subfigure}
        \begin{subfigure}[ht]{0.24\textwidth}
        \includegraphics[width=\textwidth]{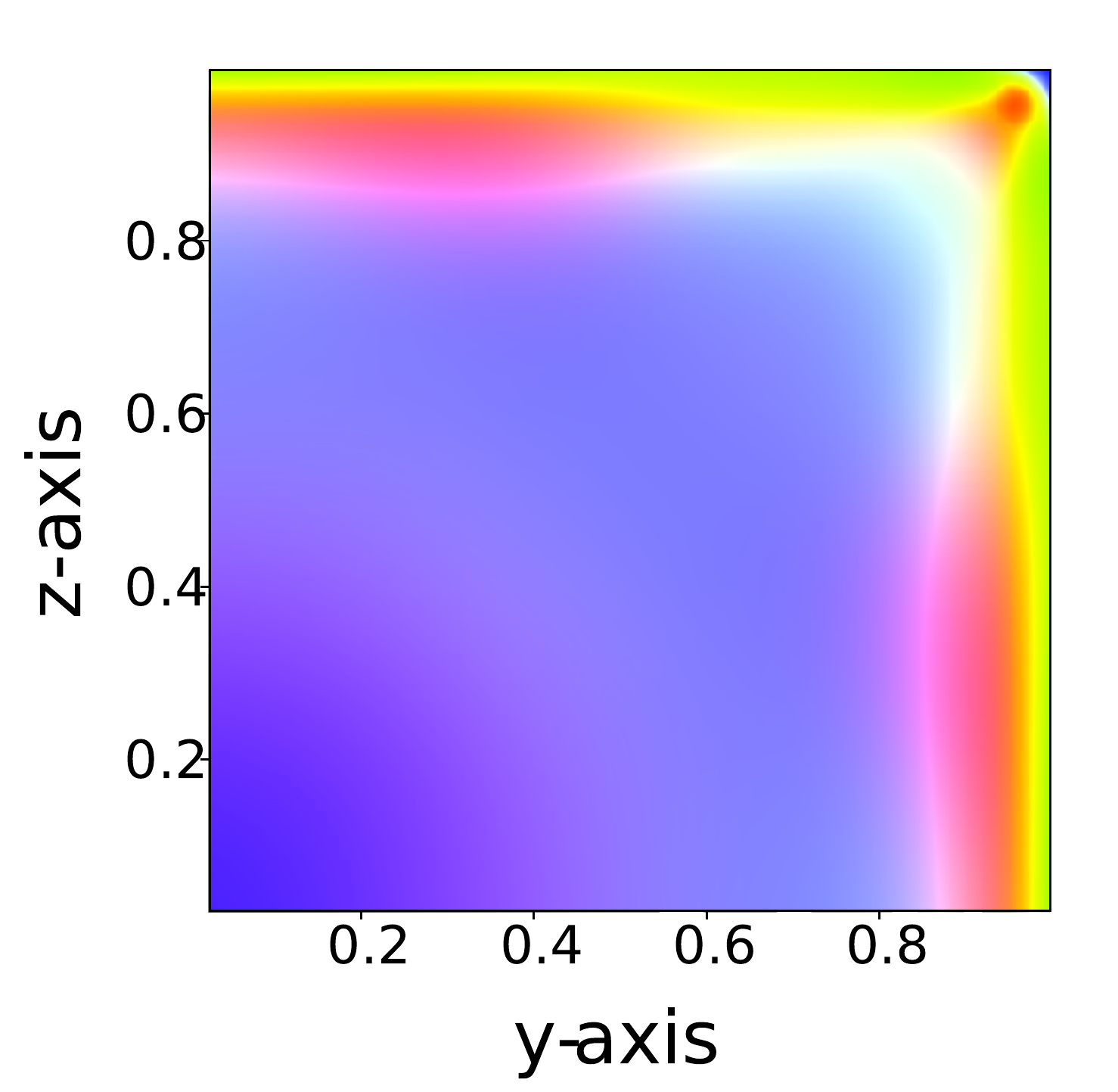}
        \caption{TBNN}
        \label{fig:InvariantMap3}
    \end{subfigure}
        \begin{subfigure}[ht]{0.24\textwidth}
        \includegraphics[width=\textwidth]{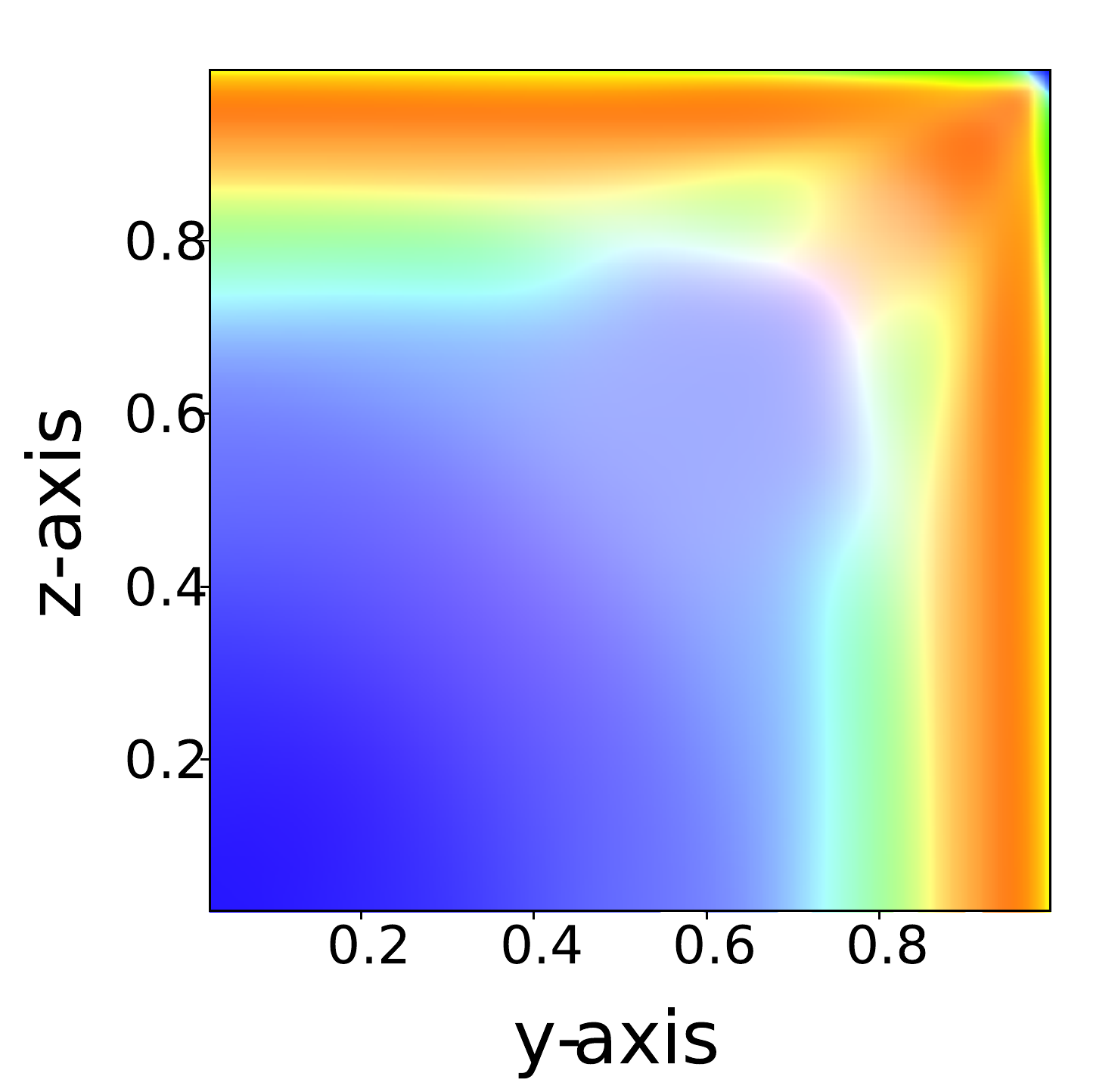}
        \caption{TBRF}
        \label{fig:InvariantMap4}
    \end{subfigure}
        \caption{Turbulence anisotropy state in the square duct (upper-right quadrant), visualized with the RGB colormap (Fig.~\ref{fig:Bary}(a)).  TBNN = Tensor-Basis Neural-Network; TBRF = Tensor-Basis Random-Forest.}
    \label{fig:InvariantMap} 
\end{figure}

One further method of visualization we will use is the Red-Green-Blue (RGB) map, in which each anisotropic state is assigned a colour and the flow domain is coloured accordingly, in a kind of generalized contour plot \citep{Tracey2013}. Figure~\ref{fig:Bary}(a) presents this colormap, and in Figure~\ref{fig:InvariantMap} the colormap is applied to the square-duct with DNS and RANS data (the same data used for Figure~\ref{fig:Bary}(b-c)).  The DNS data shows 1-component turbulence close to the wall, and 3-component near the centreline of the duct, whereas the RANS simulation only represents turbulence near the 3-component limit.  Also in this figure (c) and (d), machine-learning predictions of the same invarients are plotted, to be discussed later in Section~\ref{sec:results}.

\subsection{Invariance of tensor-valued functions}
\label{subsec:invariance}
The Navier-Stokes equations are Galilean invariant, i.e.\ unchanged by choice of inertial frame.  It is a physical requirement that any model for the anisotropy tensor also be frame invariant, and thereby satisfy the simple requirement that the functional model should not depend on the choice of coordinate system.  In fact this proves to be a critical requirement for the success of our machine-learning strategy, see Section~\ref{sec:results}.  Let $\Qf\in \Rbb^{3\times 3}$ be an arbitrary real orthogonal transformation; then a scalar-valued function $f: \Rbb^{3\times 3} \rightarrow \Rbb$, with tensor argument $\Sf\in\Rbb^{3\times 3}$ is frame invariant if and only if:
\begin{equation}
f(\Sf) = f(\mathsfbi{QSQ}^T),\quad \forall \Qf,\Sf.
\label{eq:inv_ex1}
\end{equation}
Similarly a tensor-valued function $\mathsfbi{h}: \Rbb^{3\times 3} \rightarrow \Rbb^{3\times 3}$ is frame invariant if and only if (e.g.\ \cite{Speziale1991}):
\begin{equation}
\Qf \, \mathsfbi{h}(\Sf) \Qf^T = \mathsfbi{h} (\mathsfbi{Q S Q}^T)\quad \forall \Qf,\Sf.
\label{eq:inv_ex2}
\end{equation}
One means of finding a $\mathsfbi{h}$ satisfying \eqref{eq:inv_ex2}, is to start with a scalar function $h:\Rbb\rightarrow\Rbb$, and specify that $\mathsfbi{h}$ is a natural generalization of $h$.  See \cite{Higham2008} for several standard definitions for $\mathsfbi{h}$ given $h$ (e.g. in terms of power-series).  Under all these definitions, not only do we have $\mathsfbi{h}(\Xf\Sf\Xf^{-1}) = \Xf \mathsfbi{h}(\Sf)\Xf^{-1}$ for any invertible matrix $\Xf$ (implying frame invariance by setting $\Xf \equiv \Qf$); but also other properties such as $\mathsfbi{h}(\Sf^T) = \mathsfbi{h}(\Sf)^T$ and $\mathsfbi{\lambda} = \mathrm{eigval}\{\Sf\} \implies h(\mathsfbi{\lambda}) = \mathrm{eigval}\{\mathsfbi{h}(\Sf)\}$.

In addition we assume that $\mathsfbi{h}$ has a power-series representation
\[
\mathsfbi{h}(\Sf) = \sum_{i=0}^\infty a^{(i)}(\mathsfbi{\lambda}) \Sf^i,\quad \mathsfbi{\lambda} = \mathrm{eigval}\{\Sf\},\quad a^{(i)}:\Rbb^3\rightarrow\Rbb,
\]
for some scalar-valued functions $a^{(i)}$ whose arguments are the invariants of $\Sf$.  We reduce this infinite sum to a finite sum with the following trick: by the Cayley-Hamilton theorem, every matrix satisfies its own characteristic equation $q(\Sf)=\mathsfbi{0}$.  However $q$ is a polynomial of degree $3$ (in 3d), whose coefficients are functions only of the invariants of $\Sf$.  Hence using $q$ we can recurrsively express powers $\Sf^3$ and higher in terms of $\If$, $\Sf$ and $\Sf^2$.  As a result there must exist an equivalent expression for $\mathsfbi{h}$:
\[
\mathsfbi{h}(\Sf) = \sum_{i=0}^2 \tilde a^{(i)}(\mathsfbi{\lambda}) \Sf^i,
\]
for some different scalar-valued functions $\tilde a^{(i)}:\Rbb^3\rightarrow\Rbb$.

In our application we seek a function from {\it multiple} tensors to the anisotropy tensor $\bfff$, meaning a generalization of the above is required.  The result remains -- under the above assumptions -- that the most general $\mathsfbi{h}$ can be written in terms of a finite tensor-basis known as the {\it integrity basis}.  
%

In particular when deriving nonlinear eddy-viscosity models, it is sometimes assumed that the Reynolds stresses are a function of the local, normalized, mean rates of strain $\Sf$ and rotation $\Rf$.  I.e. $\bfff := \bfff(\Sf,\Rf)$, where
\begin{equation}
\Sf = \frac{1}{2} \frac{k}{\epsilon} \left( \nabla \ub + \nabla \ub^T \right), \quad \Rf = \frac{1}{2} \frac{k}{\epsilon} \left( \nabla \ub - \nabla \ub^T \right).
\end{equation} 
In this case \citep{Pope1975} there are $10$ integrity basis tensors $\Tf^{(m)}$, making the most general expression for $\bfff$:
 \begin{equation}
\mathsfbi{b} = \mathsfbi{h}(\Sf,\Rf) = \sum_{m=1}^{10} \Tf^{(m)}(\Sf,\Rf) \; g^{(m)}(\theta_1, ... , \theta_5),
\label{eq:TensorBasis_Ling}
\end{equation}
where $g^{(m)}$ are scalar functions of the invariants $\theta_i$.  
%
The basis tensors derived from $\Sf$ and $\Rf$ are \citep{Pope1975}:
\begin{equation*}
\begin{array}{lllclll}
\Tf^{(1)} & = & \Sf & ~ & \Tf^{(6)} & = & \Rf^2\Sf+\Sf\Rf^2 - \frac{2}{3}\If \cdot \mathrm{trace}(\Sf\Rf^2) \\
\Tf^{(2)} & = & \Sf\Rf - \Rf \Sf & ~ & \Tf^{(7)} & = & \Rf \Sf\Rf^2 - \Rf^2\Sf\Rf \\
\Tf^{(3)} & = & \Sf^2 - \frac{1}{3}\If \cdot \mathrm{trace}(\Sf^2) & ~ & \Tf^{(8)} & = & \Sf\Rf \Sf^2 - \Sf^2\Rf \Sf \\
\Tf^{(4)} & = & \Rf^2 - \frac{1}{3}\If \cdot \mathrm{trace}(\Rf^2) & ~ & \Tf^{(9)} & = & \Rf^2\Sf^2 + \Sf^2\Rf^2 - \frac{2}{3}\If \cdot \mathrm{trace}(\Sf^2\Rf^2) \\
\Tf^{(5)} & = & \Rf \Sf^2 - \Sf^2\Rf & ~ & \Tf^{(10)} & = & \Rf \Sf^2\Rf^2 - \Rf^2\Sf^2\Rf \\
\end{array}
\label{eq:TB_T}
\end{equation*}
with invariants
\begin{equation*}
\theta_1 = \mathrm{trace}(\Sf^2) \quad \theta_2 = \mathrm{trace}(\Rf^2) \quad \theta_3 = \mathrm{trace}(\Sf^3) \quad \theta_4 = \mathrm{trace}(\Rf^2 \Sf) \quad \theta_5 = \mathrm{trace}(\Rf^2\Sf^2).
\label{eq:TB_inv}
\end{equation*}

In \cite{Wang2017} this approach is extended to derive a set of 47 invariants based on $\boldsymbol{\nabla \bar{u}}$, $\boldsymbol{\nabla{k}}$, and $\boldsymbol{\nabla{p}}$. This is the system we use in the following; the full feature-set will be shown in Section~\ref{subsec:features}.

\subsection{Tensor Basis Neural Network (TBNN)}
\label{subsec:TBNN}
In \cite{Ling2016a} an artificial neural network was used to represent $\mathsfbi{h}$.  By careful choice of the network topology the idea of the tensor basis is encoded into the network.  The network contains a 5-node input layer receiving the 5 scalar invariants derived from $\Sf$ and $\Rf$. These inputs then go through a densely connected feed-forward network with 10 hidden layers.  Additionally, the network contains an extra input layer consisting of the $10$ base tensors $\Tf^{(m)}$.  The output of the feed-forward network (representing the $g^{(m)}$ functions) is merged with this extra input layer, reproducing \eqref{eq:TensorBasis_Ling}.  Thereby the tensor-basis form of $\mathsfbi{h}$ and Galilean invariance is achieved. 

In this work TBNN is used as a competing method for comparison.  The implementation was obtained from the authors of \cite{Ling2016}: it is written in python using the lasagne library for building and training the neural network (source code available at github.com/tbnn/tbnn).  The same settings as there were used to aid a fair comparison.  A leaky ReLU activation function was used.  The number of hidden layers and nodes-per-layer were optimized in \cite{Ling2016}, and those values were used here.  The TBNN is trained using the Adam algorithm \citep{Kingma2014}, with the learning rate ($2.5 \times 10^{-5}$), the learning-rate decay, and the mini-batch size ($1000$) again based on \cite{Ling2016}.

Neural networks in general are challenging to train, and this was no exception.  To avoid overfitting, the data was randomly partitioned into training (80\%) and validation (20\%) sets. Early-stopping was used, which terminates training when the training error reduces, but the validation error starts consistently increasing. Since the validation error as a function of the epoch has a noisy behaviour, a moving average of five samples was taken to determine when early-stopping should be activated.  Initial network weights were randomly chosen, and the TBNN was trained five times from which the network was selected which performed best on the validation set. 

\subsection{Tensor Basis Random Forest (TBRF)}
\label{subsec:TBRF}

Decision trees base their predictions on a series of if-then tests on the input. Random forests consist of collections of decision trees with some randomized component differentiating trees.  Multiple decision tree algorithms exist, of which the CART (Classification And Regression Tree) algorithm serves as the starting point for the Tensor Basis Decision Tree (TBDT), which is used in the Tensor Basis Random Forest (TBRF). A brief overview of the TBRF algorithm is presented here, for a more technical overview the reader is referred to the appendix.

In the training phase of the CART decision tree, the feature space is recursively split into two bins.  In each bin a constant value is used to approximate the training data.  Given $p$ features let the training data consist of $\Xf\in \mathbb{R}^{p \times N}$ point locations in feature-space, and corresponding $\yf\in\Rbb^N$ scalar output values.  Each split is aligned with an input feature, and therefore the location of the split is completely specified by a splitting feature index $j\in\{1,\dots,p\}\subset\mathbb{N}$, and value $s$.  The two bins in which the data is split are denoted $R_L\subset\mathbb{R}$ (left) and $R_R\subset\mathbb{R}$ (right), and are given by
\begin{equation}
R_L(j,s) = \{\Xf \, | \, [\Xf]_j \leq s\} \quad R_R(j,s) = \{\Xf \, | \, [\Xf]_j > s\}.
\label{eq:TBDT_bins}
\end{equation}

For the TBDT, constant values are chosen for the tensor basis coefficients $g^{(m)}$ (see \eqref{eq:TensorBasis_Ling}) in each bin, which will be denoted by $g_L^{(m)}$ and $g_R^{(m)}$ for $R_L$ and $R_R$ respectively.  The values are chosen such that the mismatch with respect to the reference DNS/LES anisotropy tensor $\mathsfbi{b}$ is minimized. The cost function can be defined as:
\begin{equation}
J = \sum_{x_i \in R_L(j,s)} \left\| \sum_{m=1}^{10} \Tf_i^{(m)} g_L^{(m)} - \mathsfbi{b}_i \right\|_F^2 + \sum_{x_i \in R_R(j,s)} \left\| \sum_{m=1}^{10} \Tf_i^{(m)} g_R^{(m)} - \mathsfbi{b}_i \right\|_F^2,
\label{eq:TBDT_algor1}
\end{equation}
where the Frobenius norm is used to calculate the difference between the reference DNS/LES anisotropy tensor, and the tensor resulting from the tensor basis. It can be shown by setting the derivative of $J$ with respect to $g^{(m)}$ in each bin to zero, the optimum value for the 10 tensor basis coefficients in each bin is found using
\begin{equation}
\gf = \left(\sum_{i=1}^N \hat{\Tf}_i^T \hat{\Tf}_i \right)^{-1}\left(\sum_{i=1}^N \hat{\Tf}_i^T \hat{\boldsymbol{b}}_i\right).
\label{eq:TBDT_algor5}
\end{equation}
were $\hat{\Tf}_i$ and $\hat{\boldsymbol{b}}_i$ are matrices containing the flattened basis tensors and reference DNS/LES anisotropy tensors:
\begin{equation}
\hat{\Tf}_i = \left[ \begin{array}{cccc}
  [\Tf_i^{(1)}]_{11} & [\Tf_i^{(2)}]_{11} & \cdots & [\Tf_i^{(10)}]_{11} \\
  \lbrack\Tf_i^{(1)}\rbrack_{12} & [\Tf_i^{(2)}]_{12} & \cdots & [\Tf_i^{(10)}]_{12} \\
  \vdots  & \vdots  & \ddots & \vdots  \\
  \lbrack\Tf_i^{(1)}\rbrack_{33} & [\Tf_i^{(2)}]_{33} & \cdots & [\Tf_i^{(10)}]_{33} 
 \end{array} \right], \quad
\hat{\bfff}_i = \left[ \begin{array}{c}
 \lbrack \bfff_i\rbrack_{11} \\
 \lbrack \bfff_i\rbrack_{12}  \\
 \vdots    \\
 \lbrack \bfff_i\rbrack_{33}
\end{array}\right].
\label{eq:TBDT_algor4}
\end{equation}

In other words, during the training of the TBDT, each split in the tree is made by solving two least squares problems for $g^{(m)}$, for each $j$, and each value $s$, and selecting the combination which minimizes $J$.  The outer minimization is solved by brute-force over features $j$, and one-dimensional optimization is used over $s$.  The Brent 1d-optimization algorithm is used, offering a good trade-off between speed and robustness (by falling back on the golden section search in the worst case) \citep{Brent1973}.  When the number of samples in a bin falls below a threshold, we switch to brute-force search for $s$.  This threshold was set to 150 samples to limit the training time of the decision trees, a sensitivity study showed no significant influence of the threshold on the performance of the TBRF algorithm.  The splitting of the TBDT branches is terminated at either a specified maximum branching depth, or at a minimum number of samples per bin.  Due to the redundancy of the tensor-basis for any given sample $i\in\{1,\dots,N\}$, \eqref{eq:TBDT_algor5} can become ill-posed, especially towards the leaves of the tree, when only a few samples remain in a bin.  Therefore some $L^2$-regularization is added to $J$ with coefficient $\Gamma\in\mathbb{R}^+$, c.f.\ ridge regression, see the appendix for more details.  

In the tensor basis random forest, multiple tensor basis decision trees trained on a collection of bagged data sets are averaged.  Bagging implies data is repeatedly randomly drawn (with replacement) from the full data-set.  Bagging is expected to work well when combining a number of high-variance, low-bias estimators, such as the decision tree.  By averaging many noisy, but unbiased models, the variance of the prediction is reduced \citep{Hastie2008}. 
The variance of the predictions will be reduced most effectively if the errors in the component models are as far as possible uncorrelated.  This is encouraged by introducing some additional randomness in the individual trees: at each split, not all features, but a randomly selected subset of the available features is used for splitting.  The specific parameters used in our computations will be stated in Section~\ref{sec:results}.

Instead of taking the mean over all the tensor basis decision trees, it proved to be more successful to take the median of the trees in the random forest (as for example investigated in \cite{Roy2012}), since this removed sensitivity to outliers in the predictions, see the appendix for a comparison.  Since the random forest is a piecewise constant approximation of $\bfff$, and derivatives of $\bfff$ are needed in the N-S equation, the predictions from the TBRF are smoothed spatially with a Gaussian filter, before they are propagated through the solver to obtain a flow field (see Section~\ref{subsec:NumSetup}).  The TBRF algorithm has no explicit spatial correlation in the predictions since these are based on local features of the flow, so filtering the predictions will introduce some spatial correlation.  The filter standard deviation was set to 3 cell lengths, as this sufficiently smooths the predictions while maintaining all important features of the predicted tensor (see the appendix for more details).  This filter width is an ad hoc choice, and can possibly be adjusted more specifically for numerical stability in future work by looking at e.g. required condition numbers for the solver (see e.g. \cite{Wu2019}).

Several benefits can be identified in using the TBRF algorithm. First of all, the available data can be divided into a training data-set and validation data-set in a natural manner. Since random samples are taken from the available data with replacement until the original size of the data-set is reached, a number of samples will not be present in the training data-set for each decision tree, called out-of-bag (OoB) samples. These OoB samples can then be used to give a validation error, or OoB error. During training of the trees, this OoB error can be observed to determine when training can be stopped. Using the OoB error is similar to performing $N$-fold cross validation \citep{Hastie2008}.  Using the OoB error allows us to optimize the number of trees in the forest during training. While hyperparameters of the TBRF were tuned (see Section~\ref{sec:results}), the algorithm is robust to the choice of hyperparameters. It will work out-of-the-box without much tuning quite well, see the appendix for more details.  Compared to neural networks, the random forest algorithm is furthermore easy to train, since one does not have to worry about selecting the appropriate optimization algorithm which has its own set of hyperparameters, and its convergence. 

The TBRF algorithm presented here was implemented in python, for the source code see \cite{Kaandorp2019}.

\subsection{Choice of input features}
\label{subsec:features}
Under the modelling assumption that the Reynolds stress tensor can be well approximated using only the mean stress and rotation tensors, $\Sf$ and $\Rf$, \cite{Pope1975}, the 5 invariants of the tensor basis \eqref{eq:TensorBasis_Ling}, namely $\thetab = (\theta_1,\dots,\theta_5)$ are sufficient to describe every possible tensor function.  In the context of machine-learning, this choice of input features was made in e.g.\ \cite{Ling2016}.  However, if we relax this assumption, then it is reasonable to also use other quantities available in the mean-flow as inputs, provided they are appropriately normalized and Galilean invariant.  In particular, in the case of the square duct (see Section~\ref{subsec:FlowCases}) it was observed that due to the symmetry of the case there are only two distinct ``basis functions'' defined by $\thetab$, and these are not sufficient to accurately describe the DNS anisotropy tensor for this case.  

Therefore here we will use the full set of invariants derived from $\Sf$, $\Rf$, and $\boldsymbol{\nabla k}$ from \cite{Wang2017}. In order to use the turbulent kinetic energy gradient it is first normalized using $\sqrt{k}/\epsilon$, and then transformed to an antisymmetric tensor:
\begin{equation}
\Af_k = -\If \times \nabla k.
\label{eq:antisymm_gradk}
\end{equation}
%
%
%
Furthermore nine extra scalar features which are more physically interpretable, such as the wall-distance based Reynolds number are used, which were obtained from \cite{Wu2016a} (which were in turn based on those presented in \cite{Ling2015}).  All features which are used are presented in Table \ref{tab:features}, where feature set 1 (FS1) is based on  $\Sf$ and $\Rf$ only, feature set 2 (FS2) additionally $\Af_k$, and feature set 3 (FS3) are additionally the features from \cite{Wu2016a}.
For the features in FS3 an normalization factor is included, whereas the tensors in FS1 and FS2 are normalized using $k$ and $\epsilon$.
Note that all features in FS3 are rotationally invariant, but some (or their normalization factors) are not Galilean invariant as they include terms depending on the velocity of the flow -- the distinction is marked with a $\dagger$ in the table. 

\begin{table}
\small
\centering
\begin{tabular}{p{1.3cm} L{3cm} p{3cm} L{4.5cm}}
\textbf{Set} & \textbf{Features} & \textbf{Normalization} & \textbf{Comment} \\ \vspace{0.2cm} \\
FS1 & $\Sf^2$, $\Sf^3$, $\Rf^2$, $\Rf^2 \Sf$, $\Rf^2 \Sf^2$, $\Rf^2 \Sf \Rf \Sf^2$ & \quad - & Invariant set based on $\Sf$ and $\Rf$ \\ \vspace{0.1cm} \\
FS2 & $\mathsfbi{A_k}^2$, $\mathsfbi{A_k}^2 \Sf$, $\mathsfbi{A_k}^2 \Sf^2$, $\mathsfbi{A_k}^2 \Sf \mathsfbi{A_k} \Sf^2$, $\Rf \mathsfbi{A_k}$, $\Rf \mathsfbi{A_k} \Sf$, $\Rf \mathsfbi{A_k} \Sf^2$, $\Rf^2 \mathsfbi{A_k} \Sf^*$, $\Rf^2 \mathsfbi{A_k} \Sf^{2*}$, $\Rf^2 \Sf \mathsfbi{A_k} \Sf^{2*}$ & \quad - & Added invariants when including $\boldsymbol{\nabla k}$ \\ \vspace{0.1cm} \\
FS3 & $\frac{1}{2} (\|\Rf\|^2 - \|\Sf\|^2)$ & $\|\Sf\|^2$  & Ratio of excess rotation rate to strain rate \\
~ & $k$ ${}^\dagger$ & $\frac{1}{2}\bar{u}_i \bar{u}_i$ & Turbulence intensity \\
~ & $\min \left( \frac{\sqrt{k}d}{50\nu},2\right)$ & \quad - & Wall-distance based Reynolds number \\
~ & $\bar{u}_k \frac{\partial p}{\partial x_k}$ ${}^\dagger$ & $\sqrt{\frac{\partial p}{\partial x_j} \frac{\partial p}{\partial x_j} \bar{u}_i \bar{u}_i}$ & Pressure gradient along streamline \\
~ & $\frac{k}{\epsilon}$ & $\frac{1}{\|\Sf\|}$ & Ratio of turbulent time scale to mean strain time scale \\
~ & $\sqrt{\frac{\partial p}{\partial x_i} \frac{\partial p}{\partial x_i}}$ & $\frac{1}{2}\rho \frac{\partial}{\partial x_k} \bar{u}_k^2$ & Ratio of pressure normal stresses to shear stresses \\
~ & $\bar{u}_i \frac{\partial k}{\partial x_i}$ ${}^\dagger$ & $|\overline{u_j' u_k'} S_{jk}|$ & Ratio of convection to production of TKE \\
~ & $\|\overline{u_i' u_j'}\|$ & $k$ & Ratio of total to normal Reynolds stresses \\
~ & $ \left| \bar{u}_i \bar{u}_j \frac{\partial \bar{u}_i}{\partial x_j}\right|$ ${}^\dagger$ & $\sqrt{\bar{u}_l \bar{u}_l \bar{u}_i \frac{\partial \bar{u}_i}{\partial x_j} \bar{u}_k \frac{\partial \bar{u}_k}{\partial x_j}}$ & Non-orthogonality between velocity and its gradient 
\end{tabular}
\caption{Features used for the machine learning algorithms, obtained from \cite{Wang2017} and \cite{Wu2016a}. For features with an * all cyclic permutations of labels of anti-symmetric tensors need to be taken in account. For FS1 and FS2 the trace of the tensor quantities is taken.  Features marked with $\dagger$ are rotationally invariant but not Galilean invariant.  
}
\label{tab:features}
\end{table}


\subsection{Propagation of the predicted anisotropy tensor}
\label{subsec:NumSetup}
The open source CFD toolbox OpenFOAM was used to calculate RANS flow fields in this work. The $k-\omega$ turbulence closure model was used, together with the second-order accurate SIMPLE (Semi-Implicit Method for Pressure Linked Equation) scheme.

Simply setting the prediction of the anisotropy tensor $\bfff_\mathrm{ML}$ in the momentum equation adversely affects the numerical stability of the solver.  As already shown in \cite{Wu2019}, treating the Reynolds stress as an explicit source term in the RANS equations can lead to an ill-conditioned model. Two main strategies are used here to improve stability: (a) under-relaxing $\bfff_\mathrm{ML}$ against the Boussinesq $\bfff_\mathrm{B}:=\nu_t \mathsfbi{\hat S}$ with a relaxation parameter $\gamma\in[0,1]$, and (b) simultaneously solving a modified $k$-equation to obtain a turbulence kinetic energy corresponding to the modified anisotropy.

In detail, the incompressible Reynolds-averaged Navier-Stokes equations are 
\begin{equation}
\frac{\partial \bar{\ub}}{\partial t} + \bar{\ub}\cdot\nabla\bar{\ub} = \nabla\cdot\left[-\bar{p} + \nu \mathsfbi{\hat S} - \taub \right]
\label{eq:NS}
\end{equation}
where $\nu$ is the molecular viscosity.  The prediction $\bfff_\mathrm{ML}$ is introduced into the momentum equation, by modelling $\taub$ as
\begin{equation}
	\taub \simeq \taub_\mathrm{ML}(\gamma) := \frac{2}{3}k\If + 2k[(1-\gamma)\bfff_\mathrm{B} + \gamma \bfff_\mathrm{ML}].
	\label{eq:NS2}
\end{equation}


%
%
The blending parameter $\gamma$ starts at $0$ and is gradually increased during the simulation, i.e.\ a continuation method, see e.g.\ \cite{Knoll2004}.  A linear ramp based on the iteration count $n$ is used:
\begin{equation*}
\gamma_n = \gamma_\mathrm{max} \min \left\{1, \frac{n}{n_\mathrm{max}} \right\},
\label{eq:continuation}
\end{equation*}
where $\gamma_\mathrm{max} \geq 0.8$ is achieved in all test-cases presented here, and $n_\mathrm{max}$ is the iteration count, after which $\gamma$ is fixed.  Sufficient iterations are performed after this point to achieve solver convergence. A lower value for $\gamma$ means that the linear eddy viscosity assumption becomes more dominant, resulting in a more stable solution, but impairing the accuracy of the solved mean velocity as already noted in \cite{Wu2019}. Here, $\gamma_\mathrm{max}$ was incremented in steps of $0.1$ until the solver became unstable, yielding a value of $\gamma_\mathrm{max} = 0.8$. As this choice is ad hoc, further work related to this topic is necessary.

Furthermore, the turbulent kinetic energy in \eqref{eq:NS2} is obtained by solving a version of the $k-\omega$ $k$-equation, in which the production term is modified to be consistent with the predicted Reynolds stress in the momentum equation.  The standard production term
\begin{equation}
\mathcal{P} = - \taub : \nabla\bar\ub,
\label{eq:productionTerm}
\end{equation}
is approximated in the $k-\omega$ model by replacing $\taub$ with its Boussinesq approximation~\citep{Wilcox2008}.  Here we use the model $\taub_\mathrm{ML}$ from \eqref{eq:NS2} instead, including the blending with $\gamma$.

With these modifications, the solver converges for $\bfff$-tensors originating from DNS, TBNN and TBRF.

\section{Results}
\label{sec:results}
We compare predictions of the TBRF algorithm just described, with baseline RANS ($k-\omega$), DNS/LES references (withheld reference data), and the TBNN algorithm with the same feature sets as TBRF.  Data for training and predicting comes from five flow cases which will be discussed briefly in Section~\ref{subsec:FlowCases}. Predictions of the anisotropy tensor itself will be presented in Section~\ref{subsec:res_predictions}; corresponding mean-flow predictions are presented in Section~\ref{subsec:res_propagations}.

\subsection{Flow cases}
\label{subsec:FlowCases}
\begin{figure}
\centering
    \begin{subfigure}[ht]{0.49\textwidth}
        \includegraphics[trim={3.7cm 0cm 4.5cm 0cm},clip,width=\textwidth]{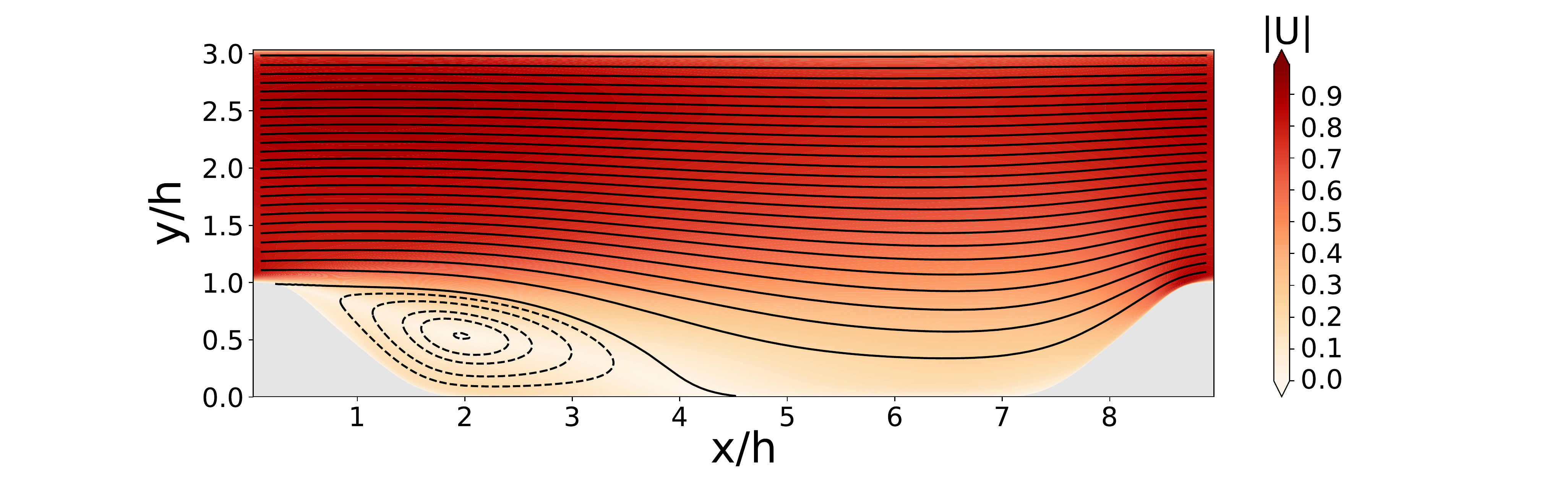}
        \caption{Periodic hills (PH)}
        \label{fig:PH}
    \end{subfigure}
    \begin{subfigure}[ht]{0.49\textwidth}
      \includegraphics[trim={3.7cm 0cm 4.5cm 0cm},clip,width=\textwidth]{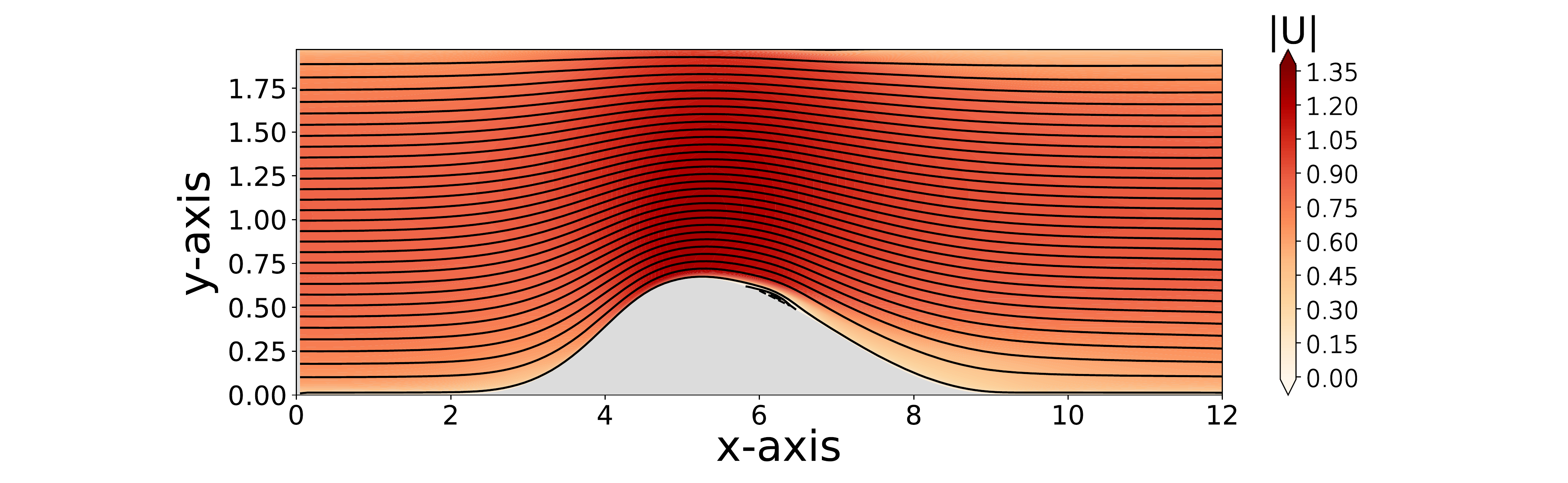}
        \caption{Converging-diverging channel (CD)}
        \label{fig:CD}
    \end{subfigure}
    \begin{subfigure}[ht]{0.49\textwidth}
        \includegraphics[trim={3.7cm 0cm 4.5cm 0cm},clip,width=\textwidth]{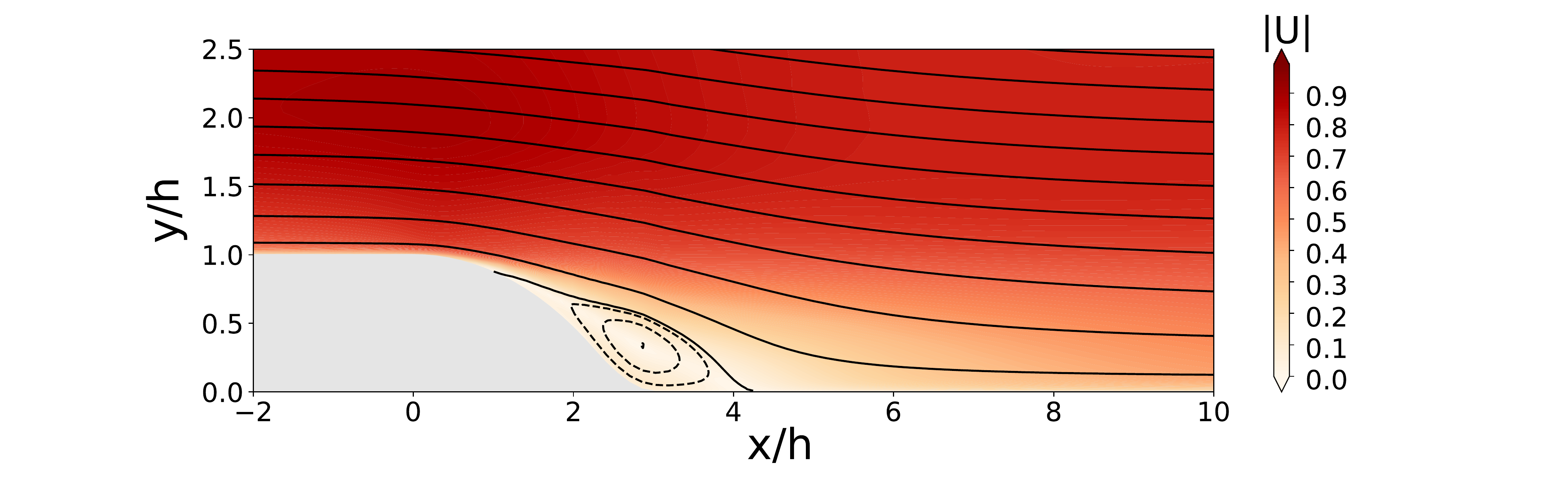}
        \caption{Curved backward facing step (CBFS)}
        \label{fig:CBFS}
    \end{subfigure}
    \begin{subfigure}[ht]{0.49\textwidth}
        \includegraphics[trim={3.7cm 0cm 4.5cm 0cm},clip,width=\textwidth]{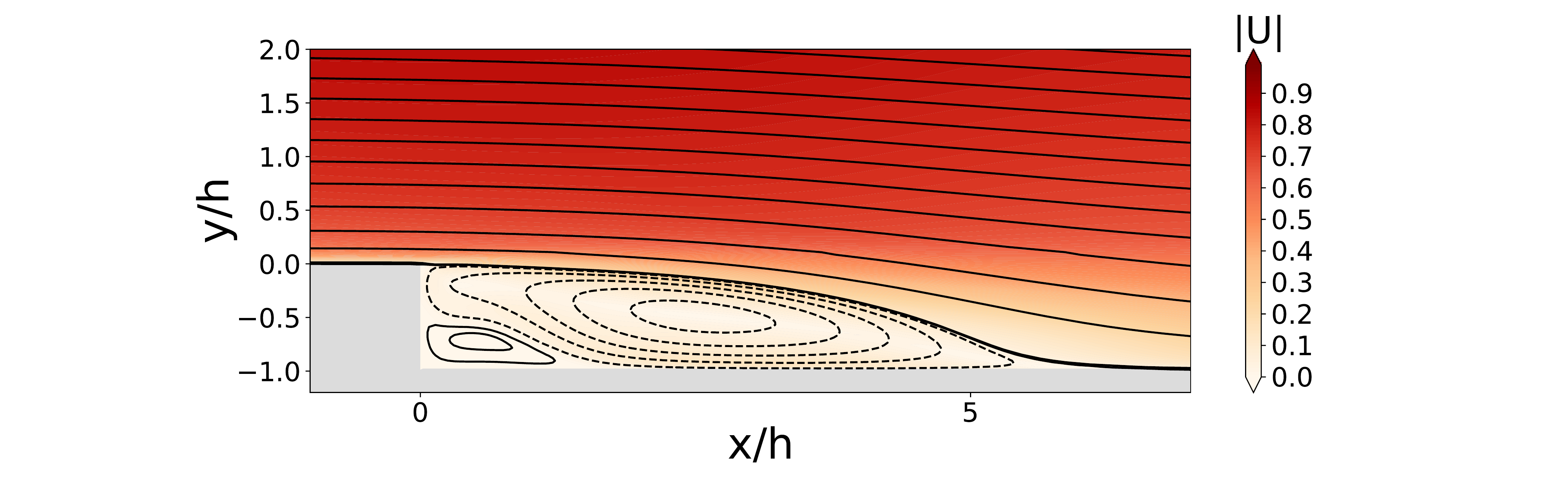}
        \caption{Backward facing step (BFS)}
        \label{fig:BFS}
    \end{subfigure}   
        \begin{subfigure}[ht]{0.4\textwidth}
        \includegraphics[width=\textwidth]{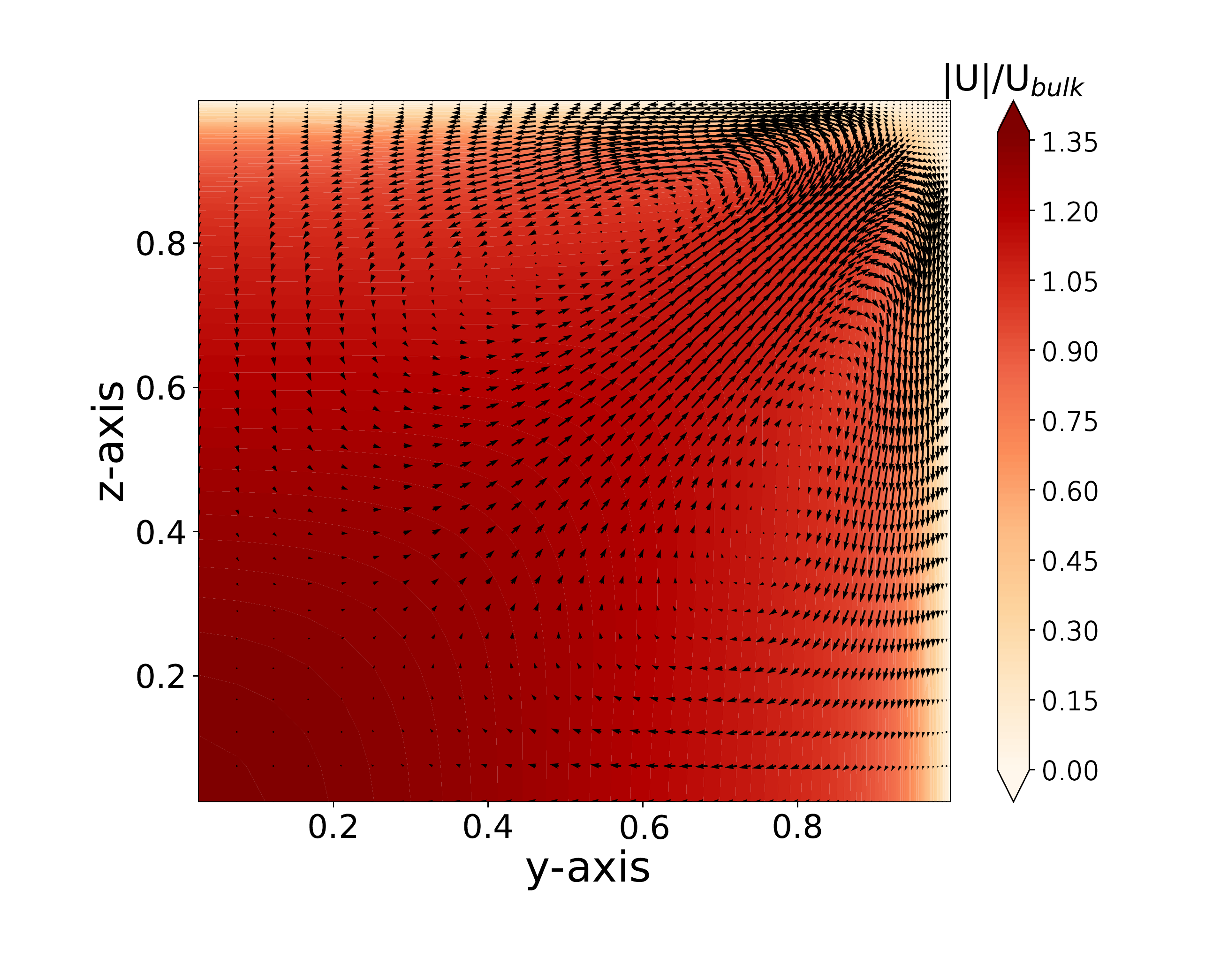}
        \caption{Square duct (SD)}
        \label{fig:SD}
    \end{subfigure}  
    \caption{Flow cases used in this work; the mean flow from the reference DNS/LES solutions is shown here. The color represents the velocity magnitude, and streamlines are plotted. Clockwise rotating regions of separation are indicated by dashed lines.}
    \label{fig:FlowCases}
\end{figure}

Five flow cases are used in the framework to train and test the machine learning algorithms. For all flow cases DNS data or highly resolved LES data was available. The mean flows for the reference DNS/LES solutions are presented in Figure~\ref{fig:FlowCases}.  The flow cases are:
\begin{enumerate}[(a) ]
	\item {\bf Periodic hills (PH)}: Five Reynolds numbers are available in the DNS/LES data-set from \cite{Breuer2009}, ranging from $Re = 700$ to $Re = 10595$ based on the bulk velocity at the inlet and the hill height.

	\item {\bf Converging-diverging channel (CD)}:  The DNS data for comes from \cite{Laval2011} at $Re = 12600$ based on the channel half-height and the maximum velocity at the inlet.

	\item {\bf Curved backward-facing step (CBFS)}: The Reynolds number available is $Re = 13700$ based on the step height and the center-channel inlet velocity, with highly resolved LES data from \cite{Bentaleb2012}. 

	\item {\bf Backward-facing step (BFS)}: $Re = 5100$ based on the step height and free stream velocity. The corresponding DNS simulation can be found in \cite{Le1997}.

	\item {\bf Square duct (SD)}: Data-sets at multiple Reynolds numbers are available from \cite{Pinelli2010}, with a total of sixteen ranging from $Re = 1100$ to $Re = 3500$ based on the duct semi-height and the bulk velocity.
\end{enumerate}

The first four aforementioned cases feature flow separation and subsequent reattachment.  Recirculation bubbles, non-parallel shear layers, and mean-flow curvature are all known to pose challenges for RANS based turbulence models.  The square duct flow case is symmetric; Figure~\ref{fig:FlowCases}(e) only presents the upper right quadrant of the duct, where the flow in the duct moves out-of-plane.  Prandtl's secondary motion of the second kind is visible, driven by turbulence anistropy.  As such it is not captured at all by linear eddy-viscosity models, which makes them ideal for isolating effects of nonlinear modelling \citep{Pecnik2007}.  For all cases mesh independence studies were performed for the RANS simulations, and meshes were chosen such that discretization error was a small fraction of the turbulence modelling error.

\subsection{Anisotropy tensor predictions}
\label{subsec:res_predictions}
In this section we examine the quality with which the Reynolds anisotropy tensor is reproduced by the TBRF.  We compare by examining (a) individual tensor components, and (b) eigenvalues in the barycentric map.

Four test-cases are presented in Table \ref{tab:trainingDataSets}, including details of training and prediction flows.  In this table the names of the flow cases have been abbreviated, and the number behind the abbreviation indicates the Reynolds number.  The table also presents the number of samples used for training, $N_\mathrm{sample}$ (randomly sampled from the total data-set), and the number of usable features present in the training sets, $N_\mathrm{feature}$.  From the available features the ones with low variance ($<1 \times 10^{-4}$) were discarded, as these either did not contain any information at all, or were largely spurious.  The starting feature sets (FS) used are those specified in Table \ref{tab:features}.  For all cases the $k-\omega$ turbulence model was used for the RANS simulations.  Hyperparameters of the TBRF were tuned for cases C1, C2, and C4 using a validation set consisting of PH2800 and SD3200. A total of 100 TBDT's were used to make the predictions.  From the $17$ available features $11$ were randomly selected for calculating each optimal split in the TBDT.  The leaf nodes were set to have a minimum of $9$ samples, and the regularization factor $\Gamma$ was set to $1 \times 10^{-12}$.  For case C3 the same setting were used, except that the trees were fully grown (i.e. each leaf node consists of one sample), and all features were used to calculate the optimal split. 

\begin{table}
\footnotesize
  \begin{center}
\def~{\hphantom{0}}
  \begin{tabular}{lccccc}
Case nr. & Training  & Prediction & $N_\mathrm{sample}$ & $N_\mathrm{feature}$ & Feature sets \\[3pt]
C1 & PH5600, PH10595, CD12600 & CBFS13700 & 21,000 & 17 & FS1, FS2, FS3 \\
C2 & PH5600, PH10595, CD12600 & BFS5100 & 21,000 & 17 & FS1, FS2, FS3 \\
C3 & PH5600, PH10595, CD12600 & SD3500 & 21,000 & 5 & FS1 \\
C4 & PH5600, PH10595, CD12600 & SD3500 & 21,000 & 17 & FS1, FS2, FS3
  \end{tabular}
  \caption{Data-sets used for training and testing. PH = Periodic Hills; CD = Converging-Diverging channel; CBFS = Curved Backward Facing Step; BFS = Backward Facing Step; and SD = Square Duct.}
  \label{tab:trainingDataSets}
  \end{center}
\end{table}
 
First prediction for the curved backward facing step will be presented (C1), which is relatively similar to the training cases for which reliable data was available (periodic hills and the converging-diverging channel).  Next, results for the backward facing step will be presented (C2), which features stronger separation than to the training cases and will therefore feature more extrapolation. Lastly, results for the square duct will be presented. A comparison will be made for the case using only features based on $\Sf$ and $\Rf$ as was also done in \cite{Ling2016} (C3), and a case where all available features are used (C4).
 
\subsubsection{Curved Backward Facing Step}
\begin{figure}
\centering
    \includegraphics[trim={3cm 0cm 3cm 0cm},clip,width=1.0\textwidth]{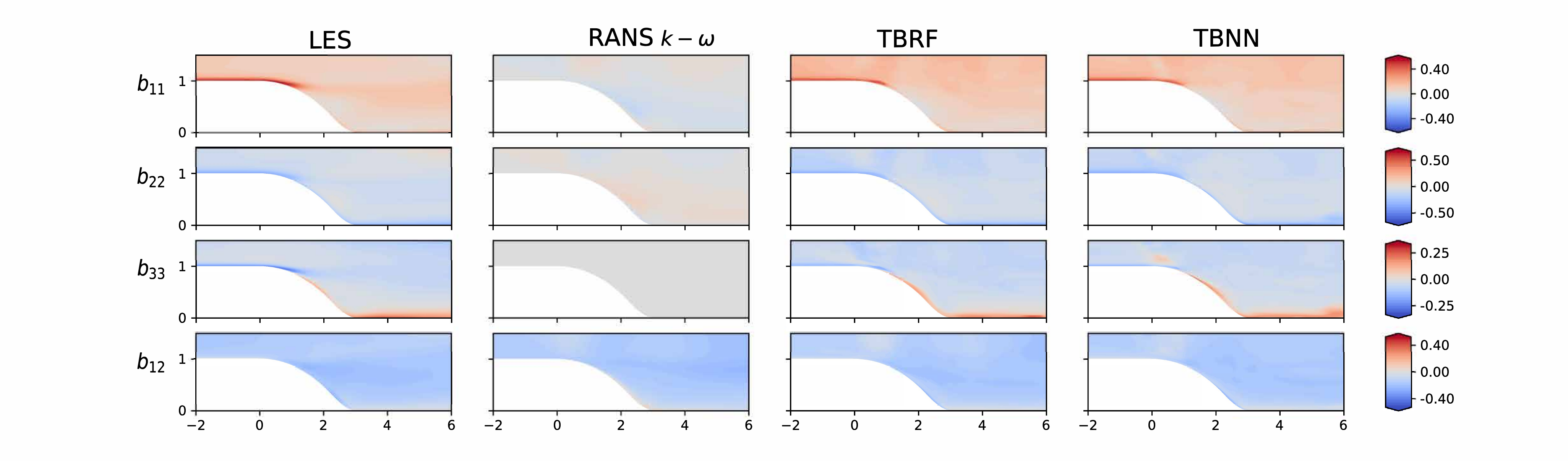}  
    \caption{Curved backward-facing step, $[\bfff]_{ij}$ components from LES, RANS, TBRF, and TBNN.}
    \label{fig:res_CBFS_components}
\end{figure}

For the curved backward facing step (case C1 in Table~\ref{tab:trainingDataSets}),
Figure~\ref{fig:res_CBFS_components} presents the four non-zero unique components of $\bfff$, as given by the LES data, the ($k-\omega$), and the TBRF and TBNN algorithms.  Taking LES as a reference, RANS only gives acceptable predictions for the $[\bfff]_{12}$ component.  By the Boussinesq assumption, results will only be acceptable where the turbulence anisotropy tensor is aligned with the mean rate of strain tensor, which is empirically not a good assumption in the majority of the domain.  In contrast, both machine learning algorithms give reasonable predictions of the tensor field in all components, and predictions are relatively smooth.  In particular, features on top of the step the $[\bfff]_{11}$ component is captured qualitatively by the TBRF and TBNN algorithms, as well as the $[\bfff]_{33}$ component after the step.
%
%
%

\begin{figure}
\centering
    \includegraphics[trim={0cm 0.0cm 0cm 0cm},clip,width=\textwidth]{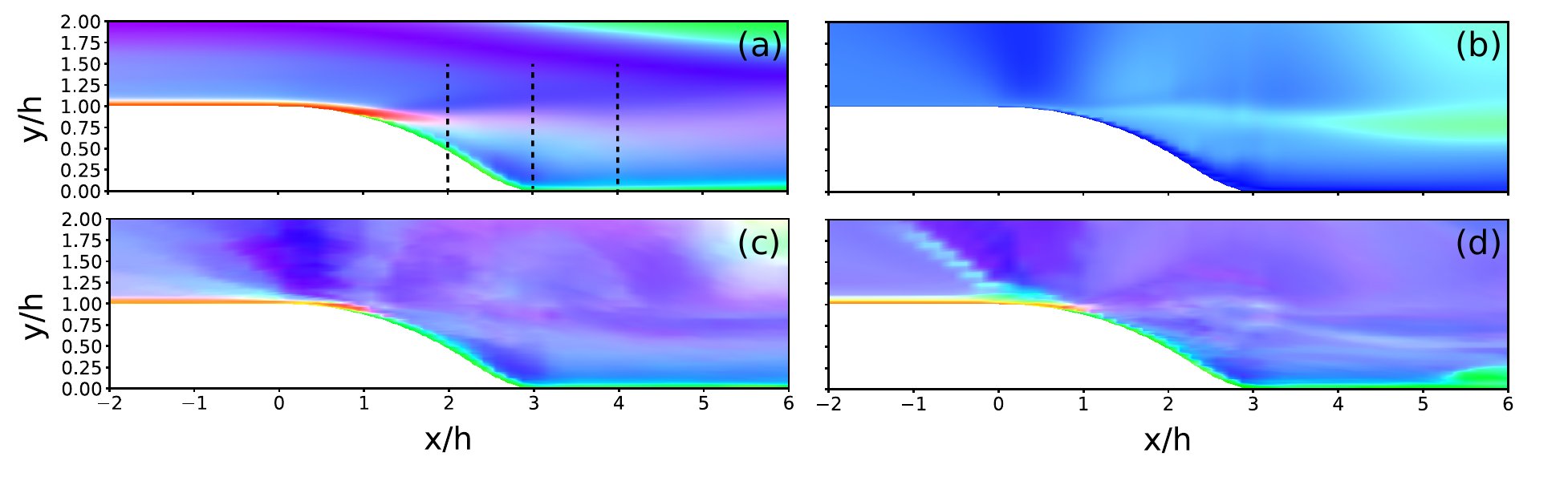}  
    \caption{Stress type for the curved backward-facing step, visualized with the RGB colormap (Fig.~\ref{fig:Bary}(a)): (a) LES from \cite{Bentaleb2012}; (b) RANS $k-\omega$, (c) TBRF, and (d) TBNN.}
    \label{fig:res_CBFS_stressType}
\end{figure}

More revealing is a plot of the stress types in the domain using the RGB map,
Figure~\ref{fig:res_CBFS_stressType}.  The LES data shows 1-component turbulence on top
of the step, which is transported into the flow, beyond the location at which the shear layer
separates. In \cite{Bentaleb2012} it is noted that production of the streamwise
fluctuation is strongly increased at the shear-layer separation location, leading to additional
1-component turbulence.  As this shear layer gains distance from the wall, the
turbulence rapidly evolves towards the 3-component state due to the redistribution
process. On the curved part of the step and the bottom wall, 2-component turbulence can
be observed.  In the remainder of the domain, 3-component turbulence dominates, in the
interior of the channel, and in the center of the recirculation region.

The RANS simulation is only capable of predicting plane-strain (c.f.\ Figure~\ref{fig:InvariantMap}),
and predicts turbulence mainly in the 3-component region.  The effect of the walls on the turbulence anisotropy is completely missed.  In contrast, the TBRF algorithm accurately captures the turbulent state as given by the LES data: 1-component turbulence can be seen on top of the hill and at the separation location, it accurately predicts the 2-component state on the curved part of the walls and on the bottom wall after the step, and 3-component turbulence can be observed in the recirculation region. Some noise is visible however, most notably around $x/h = 0.0$ to $x/h = 1.0$ away from the wall. The TBNN algorithm captures the different types of turbulence accurately as well. Close to the wall on top of the step it captures the 1-component turbulence a bit less well than the TBRF algorithm, and some spurious patterns are visible above the step and close to the lower wall around $x/h = 6.0$.

\begin{figure}
\centering
    \begin{subfigure}[ht]{0.32\textwidth}
        \includegraphics[trim={2cm 0cm 1.8cm 1.5cm},clip,width=\textwidth]{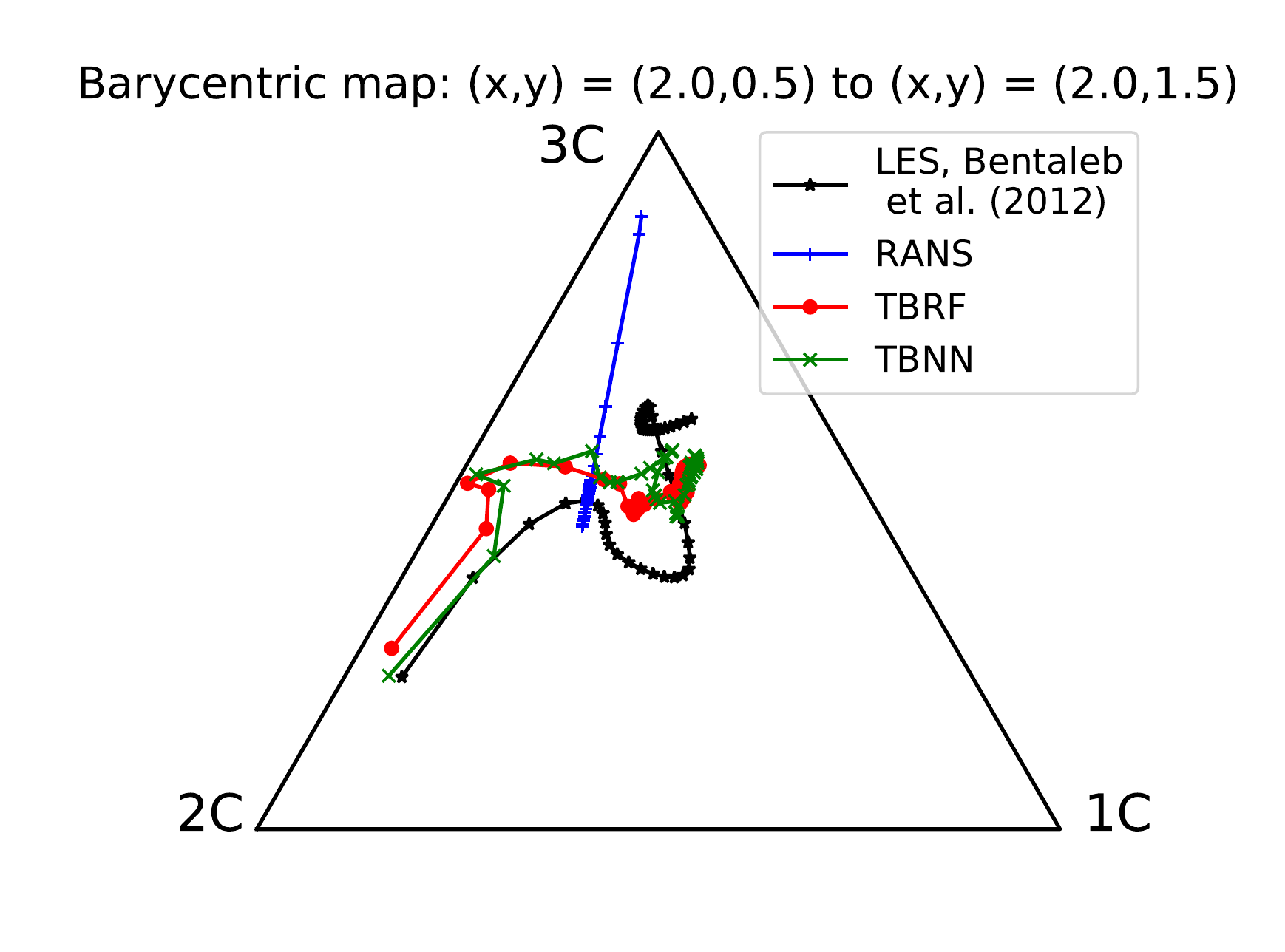}
        \caption{$x/h = 2.0$}
        \label{fig:barySection_CBFS_1}
    \end{subfigure}
    \begin{subfigure}[ht]{0.32\textwidth}
		\includegraphics[trim={2cm 0.15cm 1.8cm 1.5cm},clip,width=\textwidth]{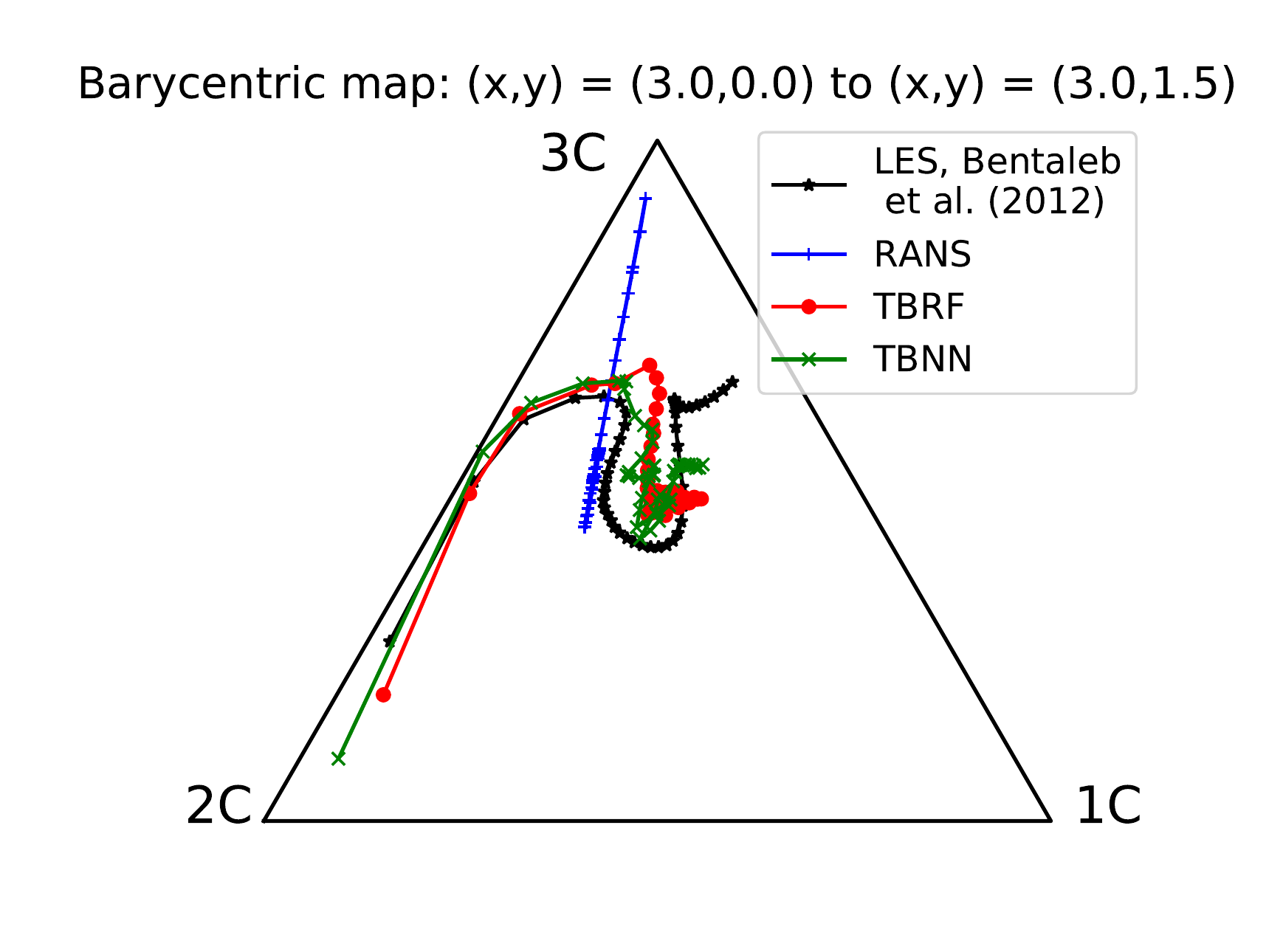}
        \caption{$x/h = 3.0$}
        \label{fig:barySection_CBFS_2}
    \end{subfigure} 
    \begin{subfigure}[ht]{0.32\textwidth}
		\includegraphics[trim={2cm 0cm 1.8cm 1.5cm},clip,width=\textwidth]{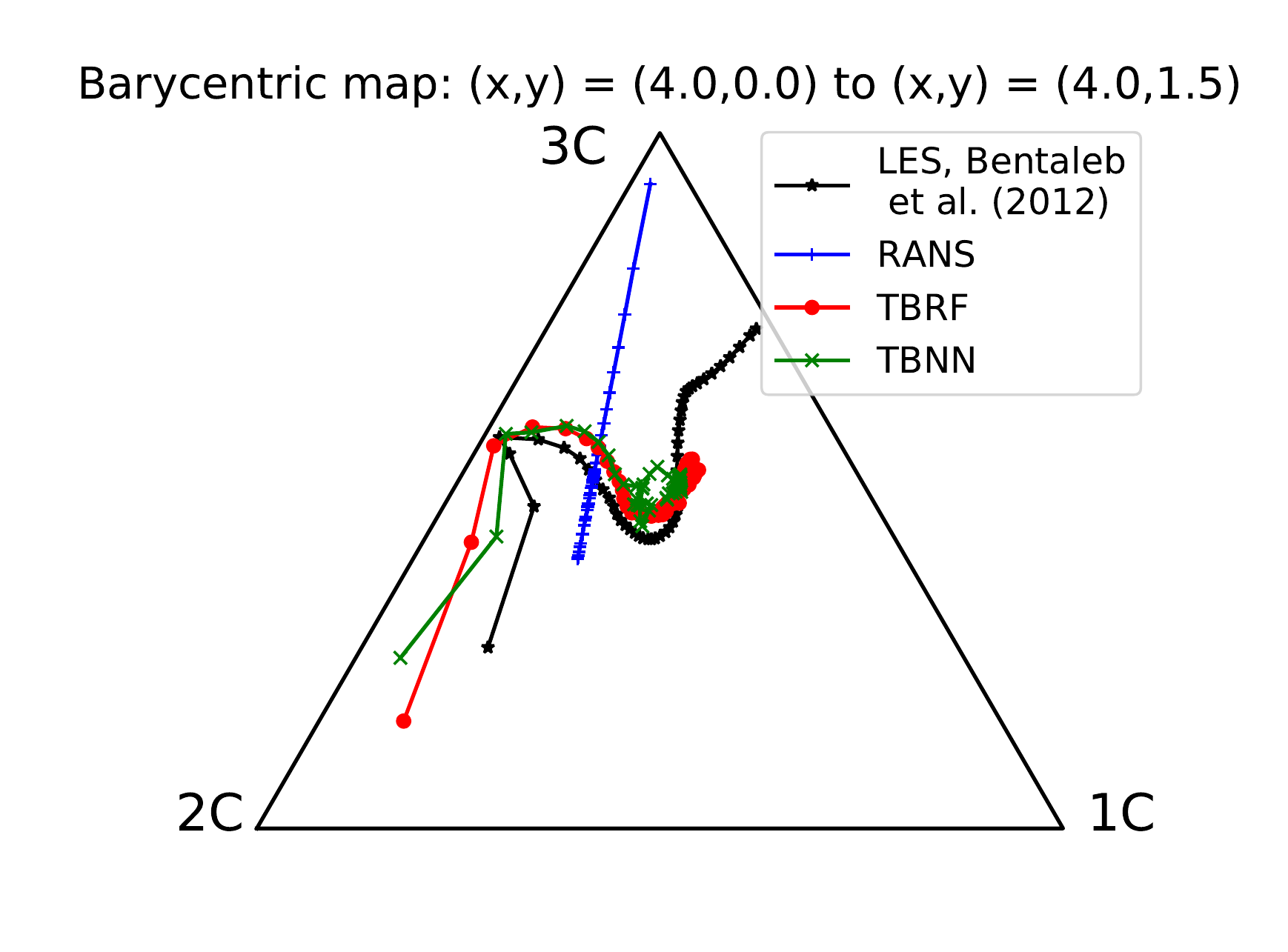}
        \caption{$x/h = 4.0$}
        \label{fig:barySection_CBFS_3}
    \end{subfigure}     
    \caption{Barycentric map data at three sections (see Figure \ref{fig:res_CBFS_stressType}a) for the curved backward-facing step.  Comparing LES \cite{Bentaleb2012}, RANS $k-\omega$ simulation, TBRF, and TBNN.}
    \label{fig:res_CBFS_barySection} 
\end{figure}

To better quantify the accuracy of reconstruction, three sections through the flow domain are plotted in the barycentric map. These sections are located at $x/h = 2$, $x/h = 3$, and $x/h = 4$, which which are at the front, middle, and aft part of the separated region.  The first section at $x/h = 2$ ranges from $y/h = 0.5$ to $y/h = 1.5$, the other two sections range from $y/h = 0.0$ to $y/h = 1.5$.  Results are presented in Figure~\ref{fig:res_CBFS_barySection}.  As can be seen both machine learning algorithms reproduce quite closely the LES reference data. They accurately capture the 2-component turbulence close to the wall, and move towards the 3-component corner when moving away from the wall. Discrepancies can be seen when moving upwards past to the wake to the channel center, where the LES data indicates a move towards axisymmetric expansion, which is less strongly represented by the ML algorithms.

\subsubsection{Backward Facing Step}
\begin{figure}
\centering
    \begin{subfigure}[ht]{0.32\textwidth}
        \includegraphics[trim={4.3cm 0cm 1.5cm 1.9cm},clip,width=\textwidth]{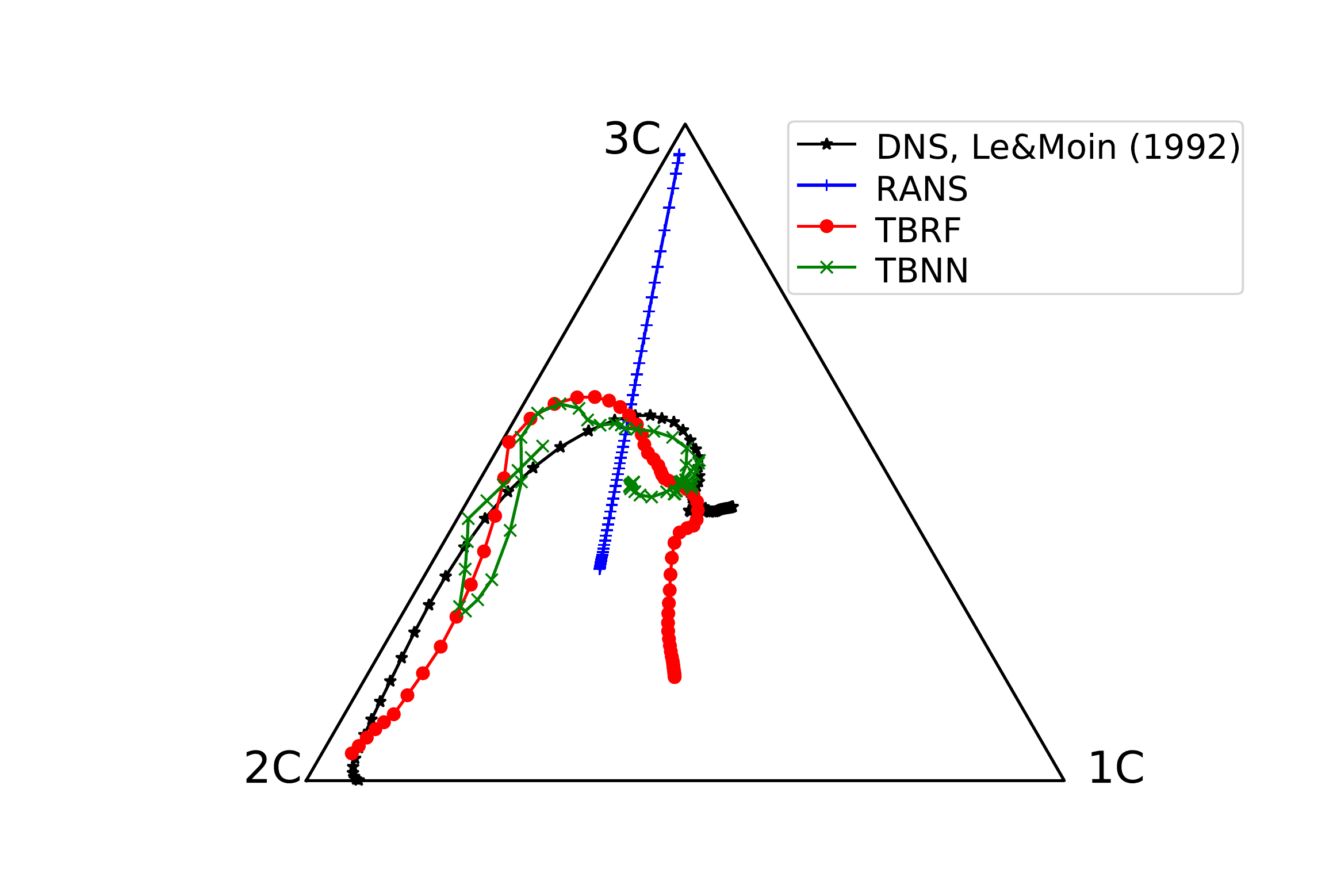}
        \caption{$x/h = 4.0$}
        \label{fig:barySection_BFS_1}
    \end{subfigure}
    \begin{subfigure}[ht]{0.32\textwidth}
        \includegraphics[trim={4.3cm 0cm 1.5cm 1.9cm},clip,width=\textwidth]{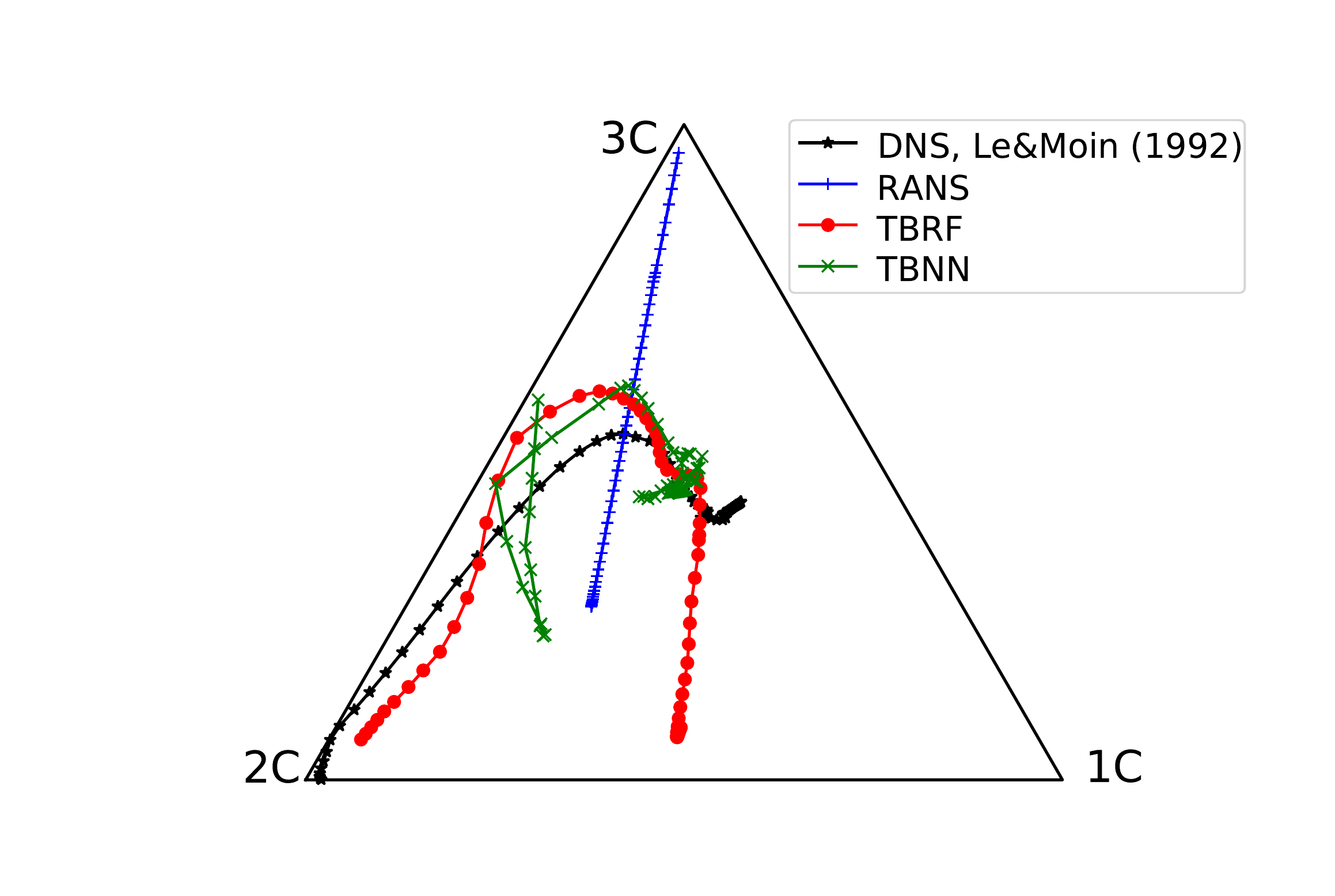}
        \caption{$x/h = 6.0$}
        \label{fig:barySection_BFS_2}
    \end{subfigure} 
    \begin{subfigure}[ht]{0.34\textwidth}
        \includegraphics[trim={4.3cm 1.3cm 1.5cm 1.9cm},clip,width=\textwidth]{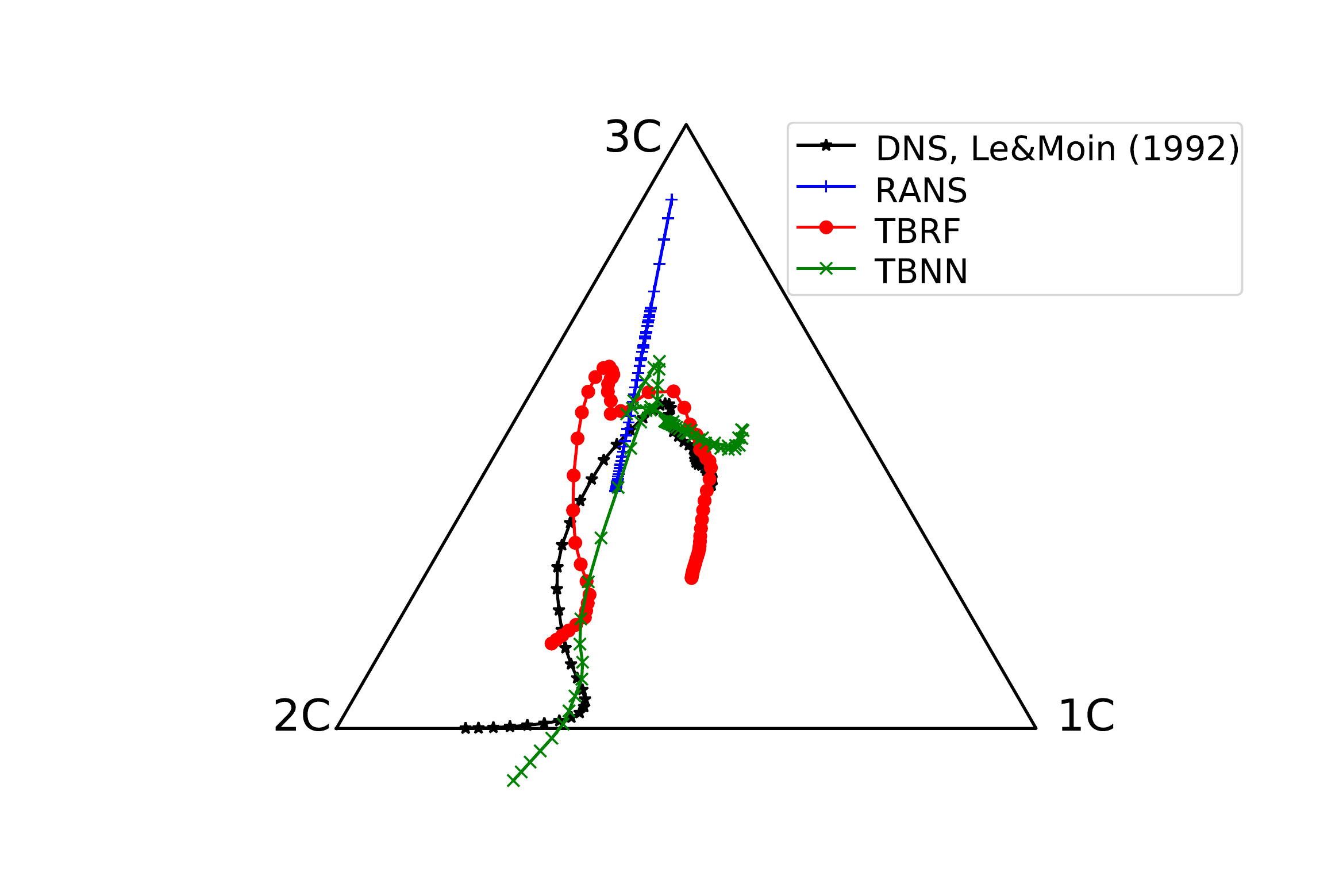}
        \caption{$x/h = 10.0$}
        \label{fig:barySection_BFS_3}
    \end{subfigure}  
    \begin{subfigure}[ht]{0.34\textwidth}
        \includegraphics[trim={4.3cm 1.3cm 1.5cm 1.9cm},clip,width=\textwidth]{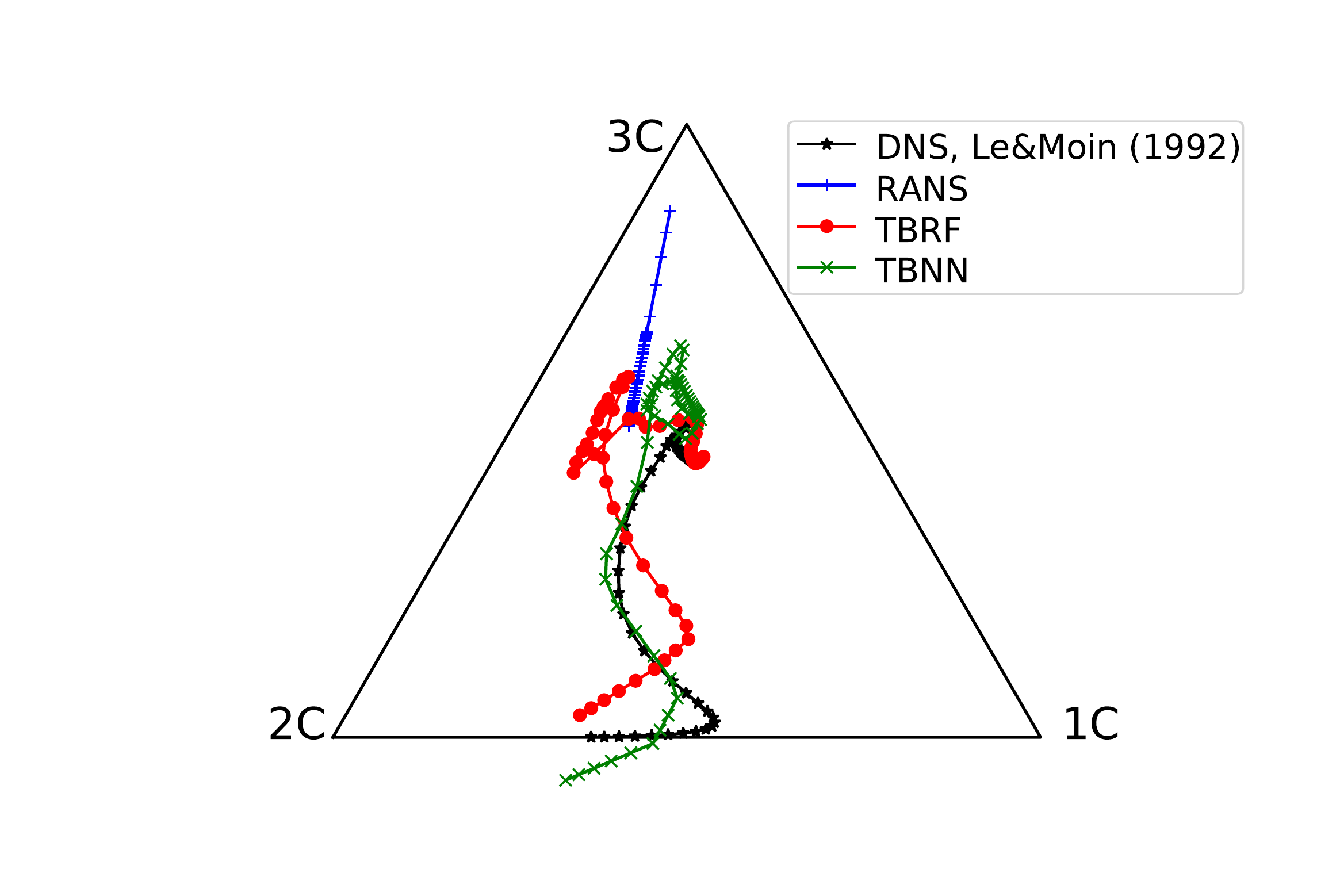}
        \caption{$x/h = 15.0$}
        \label{fig:barySection_BFS_4}
    \end{subfigure}    
    \begin{subfigure}[ht]{0.34\textwidth}
        \includegraphics[trim={4.3cm 1.3cm 1.5cm 1.9cm},clip,width=\textwidth]{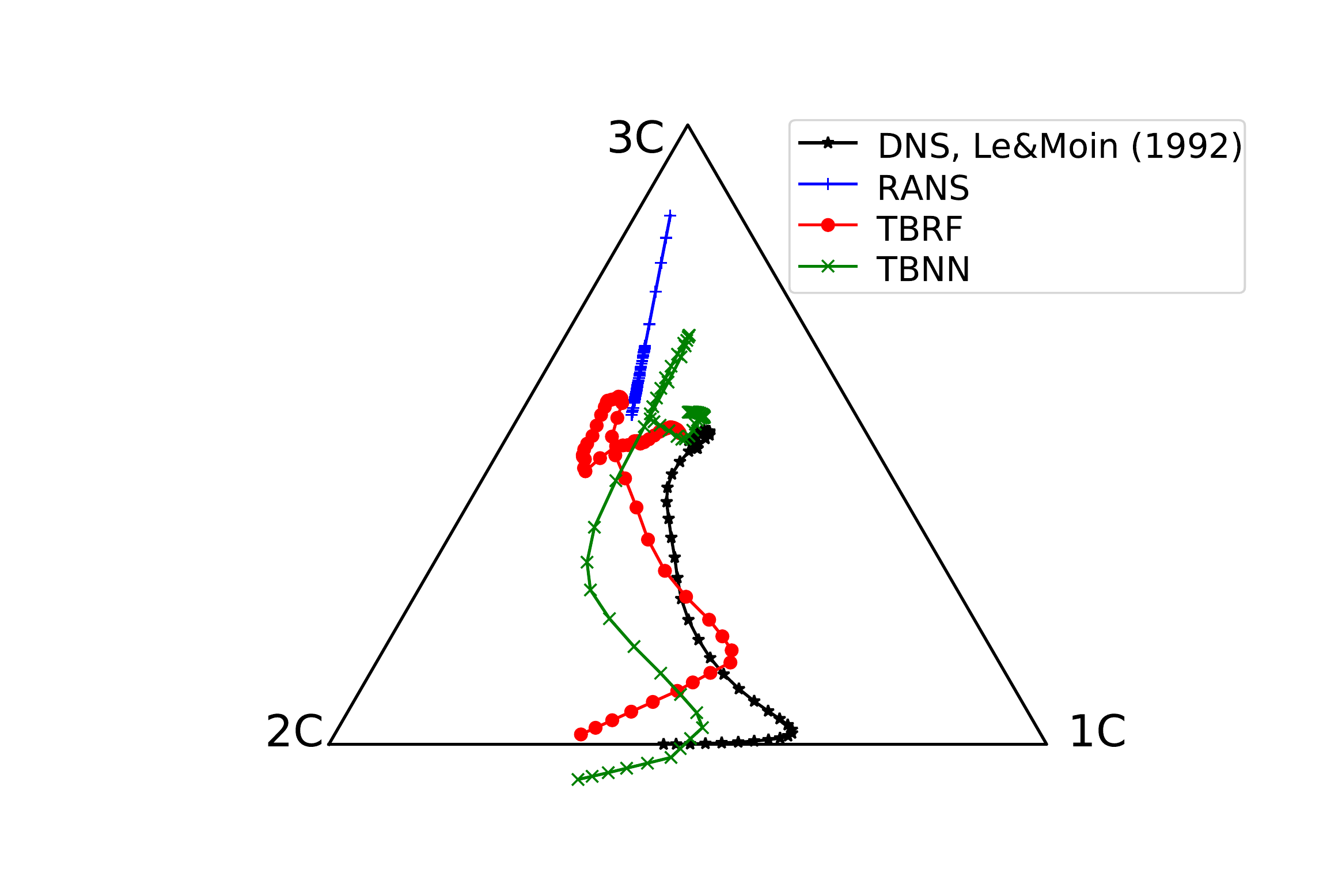}
        \caption{$x/h = 19.0$}
        \label{fig:barySection_BFS_5}
    \end{subfigure}         
    \caption{Barycentric map data at five sections for the backward-facing step. Comparing LES \cite{Le1992}, RANS $k-\omega$, TBRF, and TBNN.}
    \label{fig:res_BFS_barySection} 
\end{figure}

We consider case C2 (c.f.\ Table~\ref{tab:trainingDataSets}).  From \cite{Le1992} DNS data is available for five different sections at specified $x/h$ locations for the Reynolds stresses and velocities ($h$ is the step-height). Locations on the barycentric map for these five sections are plotted in Figure~\ref{fig:res_BFS_barySection}.  The sections range from $y/h=-1$ (bottom wall) to $y/h = 0$ (the location of the step). 

Results are similar to those of the CBFS: the machine-learning algorithms are able to give a qualitatively accurate prediction of the turbulence character, with some quantitative discrepancies.  For $x/h = 4$ and $x/h = 6$ predictions close to the wall are more accurate for TBRF than TBNN.  The situation is reversed for $x/h = 10, 15, 19$, where TBNN slightly outperforms, at the cost of some unrealizable predictions closest to the wall.  In all our studies, we have never observed unrealizable predictions from TBRF, despite no explicit realizability constraint being imposed on the method.

Moving away from the wall into the shear layer TBRF erroneously heads too far back towards the two-component boundary at the sections closest to the step.  The reason for this is unclear, at similar (shear-layer) locations in the training flows, the turbulence does not exhibit such behaviour.  Furthermore TBNN is reasonably accurate here.  Diagnostic tools are needed, and will be a focus of future research.  Nonetheless at the section further from the step, both ML methods perform well.
%

\subsubsection{Square Duct}
\begin{figure}
\centering
    \includegraphics[trim={0cm 2cm 0cm 1.5cm},clip,width=0.75\textwidth]{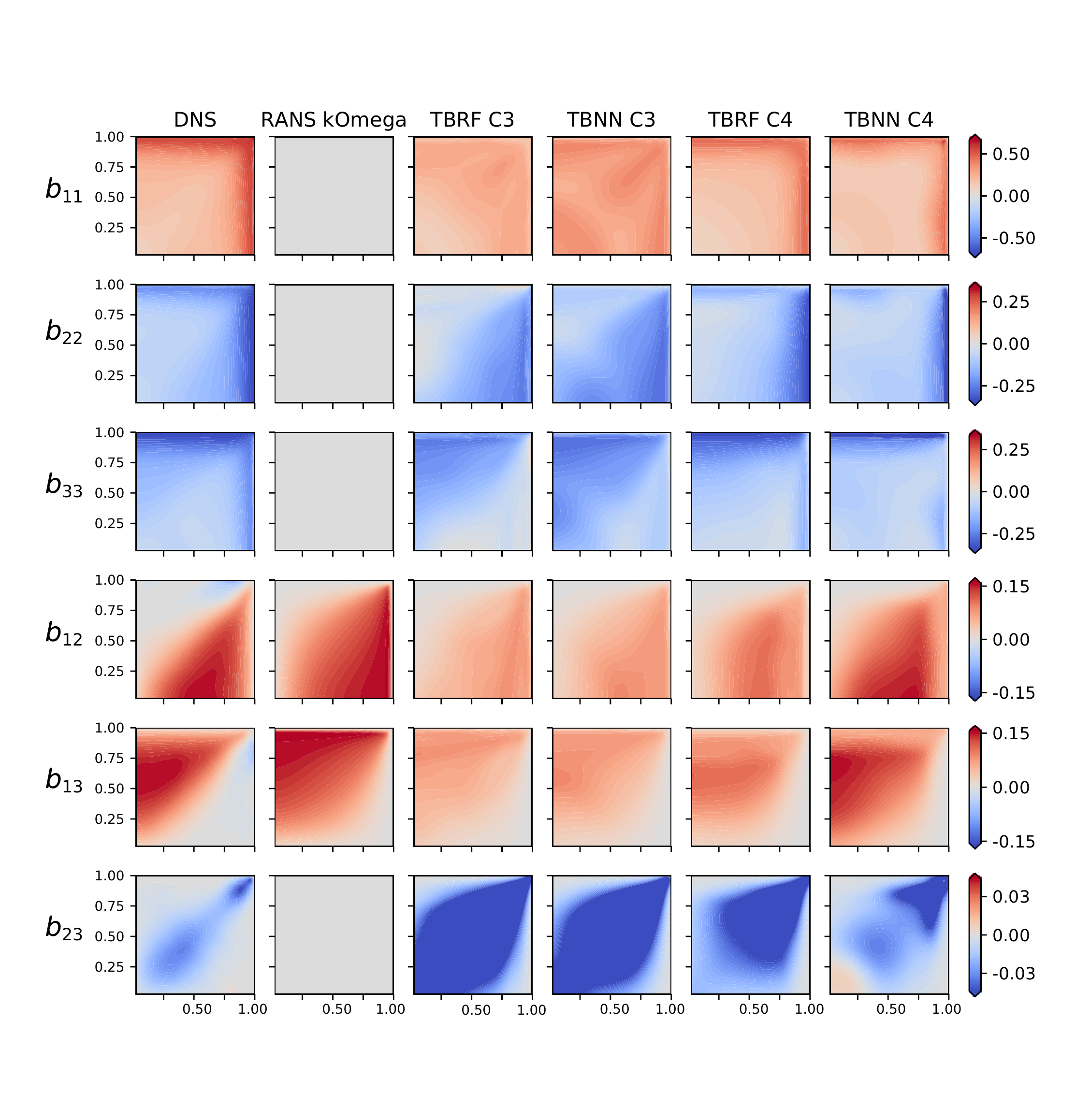}  
    \caption{Square duct, $[\bfff]_{ij}$ components from DNS, RANS, TBRF, and TBNN.}
    \label{fig:res_SD_components}
\end{figure}

The local stress type for the square duct was already shown in Figure~\ref{fig:InvariantMap}; individual components of the anisotropy tensor shown in Figure~\ref{fig:res_SD_components}.  In both cases anistropy is visualized for DNS, RANS ($k-\omega$), TBRF and TBNN predictions.  In Figure~\ref{fig:InvariantMap} ML results are only shown for case C4 (17 features); in Figure~\ref{fig:InvariantMap} additionally case C3 is shown.  Note that these are challenging cases due to the substantial differences between the training and prediction flows.

As expected, the Boussinesq model yields non-zero predictions only for $[\bfff]_{12}$ and $[\bfff]_{13}$ -- though these are relatively well predicted.  Anistropy is confined to the 3-component corner, on the plane-strain line.  Examining the predictions of ML, it can generally seen that the introduction of extra features has significantly more effect than the choice of neural-networks versus random-forests.  For example, looking at $[\bfff]_{11}$, the anisotropy of the Reynolds stress is not captured close to the walls for C3, whereas it is present in C4.  The magnitude of $[\bfff]_{12}$ and $[\bfff]_{13}$ is underpredicted, independently of the ML method, but improved in case C4 compared to C3.  Similarly in all cases the magnitude of $[\bfff]_{23}$ is over-predicted by ML, but the magnitude of the over-prediction is less in case C4.  To quantify these observations, the root mean square error (RMSE) of the anisotropy tensor with respect to the DNS data is given in Table \ref{tab:SD_RMSE}. The RMSE's are lower when introducing more features, for both algorithms.  This can largely be explained by visualizing the shapes of the input features in case C3.  Doing this it can be observed that, of the 5 features, 3 are approximately scaled versions of the other 2 -- effectively reducing the input space to two-dimensions.  This explains the difficulty nonlinear eddy-viscosity models based on only $\Sf$ and $\Rf$, have in reproducing the magnitude of the secondary flow in the square duct.
%

\begin{table}
  \begin{center}
\def~{\hphantom{0}}
\begin{tabular}{lcc}
  \hline
  Case & TBRF & TBNN  \\
  \hline
C3 & 0.0995  & 0.0871\\ 
C4 & 0.0521  & 0.0681\\
  \hline
  \end{tabular}
  \caption{RMSE of TBRF and TBNN $[\bfff]_{ij}$ predictions, for the square duct flow case.}
  \label{tab:SD_RMSE}
  \end{center}
\end{table}

\subsection{Anisotropy tensor propagation}
\label{subsec:res_propagations}
This section presents flow fields obtained by propagating the predicted anisotropy tensors for the square duct flow and the backward facing step.  Predictions for the anisotropy tensor were propagated using the stabilized solver presented in Section~\ref{subsec:NumSetup}. 

\subsubsection{Square Duct}
\begin{figure}
\centering
    \begin{subfigure}[ht]{0.48\textwidth}
        \includegraphics[trim={0.5cm 0cm 0.5cm 2cm},clip,width=\textwidth]{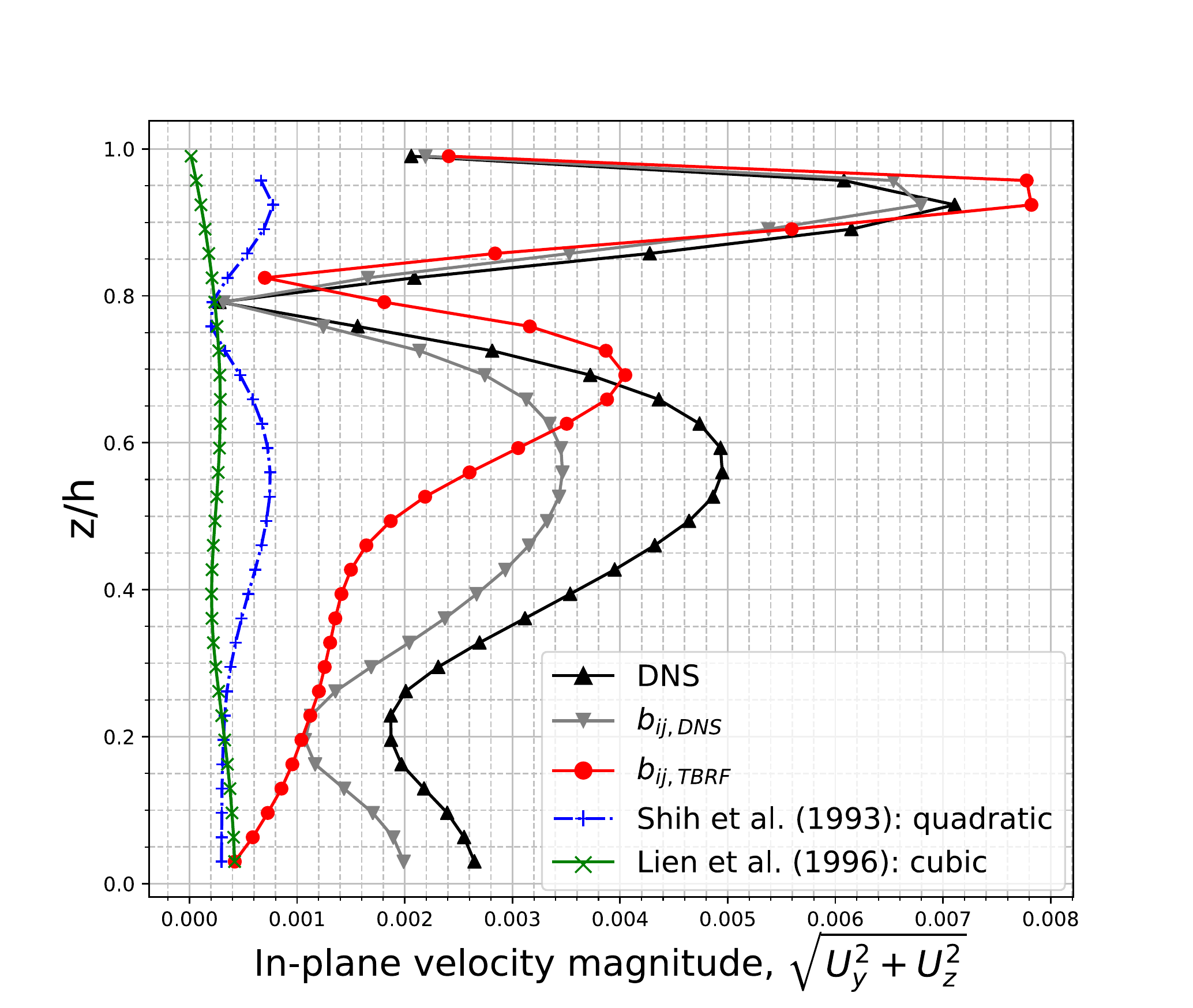}
        \caption{$y/h = 0.5$}
        \label{fig:res_propSD_section1}
    \end{subfigure}
    \begin{subfigure}[ht]{0.48\textwidth}
        \includegraphics[trim={0.5cm 0cm 0.5cm 2cm},clip,width=\textwidth]{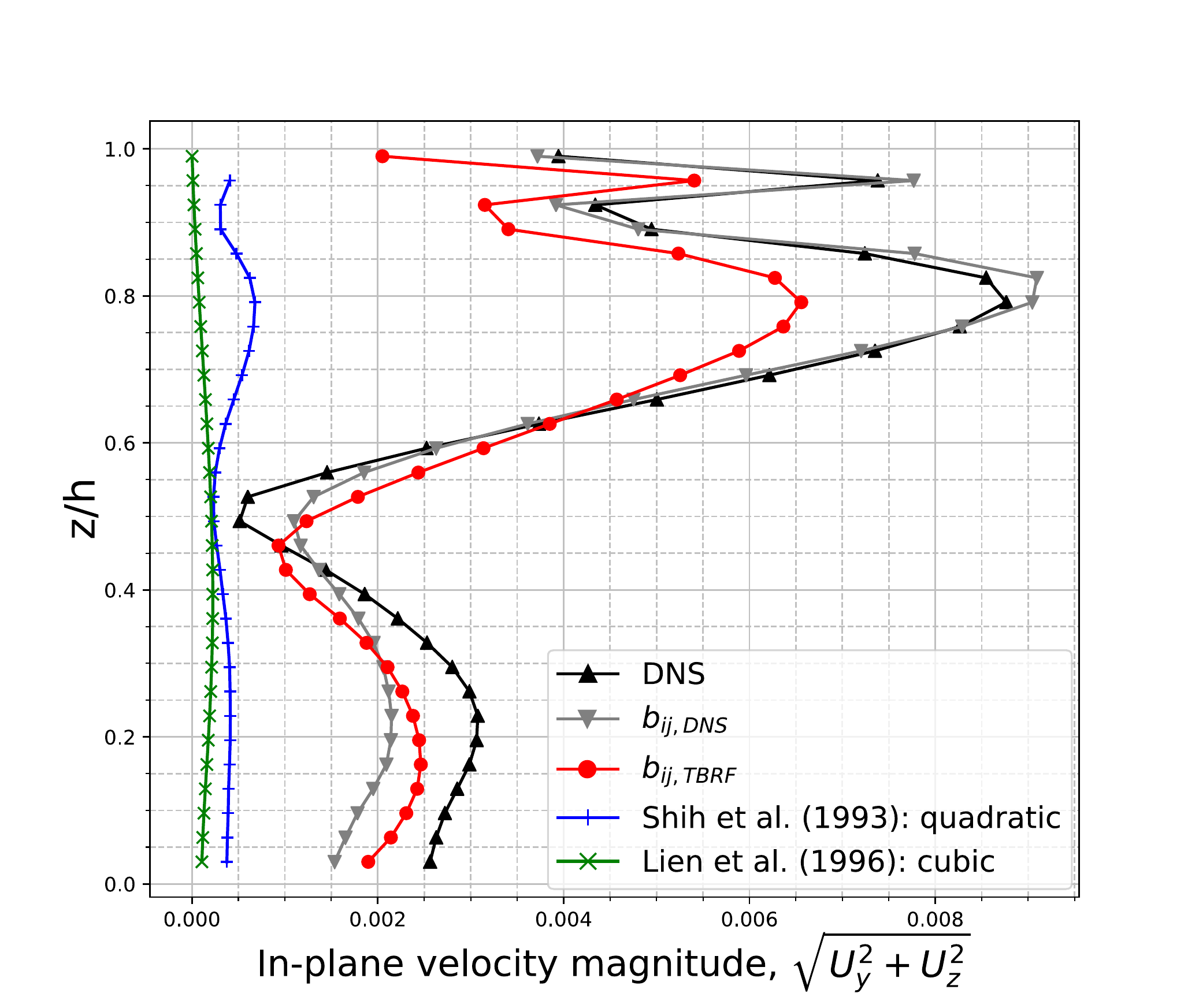}
        \caption{$y/h = 0.8$}
        \label{fig:res_propSD_section2}
    \end{subfigure}    
    \caption{In-plane mean velocity profiles at two sections of the square duct.  Comparing: DNS, and the mean velocity fields obtained by propagating the DNS anistropy, and the TBRF-predicted anisotropy (labeled $b_{ij,DNS}$ and $b_{ij,TBRF}$ respectively).  Also shown are two non-linear eddy-viscosity models.}
    \label{fig:res_propSD_section} 
\end{figure}

Two sections in the square duct will be analyzed with respect to the in-plane mean velocity magnitude (i.e. indicating the magnitude of the secondary flow). Figure~\ref{fig:res_propSD_section} presents the velocity magnitude for sections located at $y/h = 0.5$ and $y/h = 0.8$.  DNS from \cite{Pinelli2010} is used as a reference.  In order to verify the propagation method in isolation (without predicting anistropy), the anisotropy tensor obtained directly from DNS ($b_{ij,\mathrm{DNS}}$) is propagated, see the gray lines.  This is a ``best case scenario'' where the anisotropy tensor is assumed to be perfect (up to to statistical convergence of the DNS).  The mean-flow field as obtained by propagating the predictions from the TBRF algorithm ($b_{ij,\mathrm{TBRF}}$, see column 5 of Figure~\ref{fig:res_SD_components}) is indicated by the red lines.  Furthermore, results from the quadratic eddy viscosity model of \cite{Shih1993}, and the cubic eddy viscosity model of \cite{Lien1996} are presented.  Since the linear eddy-viscosity model does not yield any secondary flow at all, this result is omitted.


When examining the results of Figure~\ref{fig:res_propSD_section}, the story is broadly the same for $y/h = 0.5$ and $y/h = 0.8$.  Mean-velocity fields obtained using $b_{ij,\mathrm{DNS}}$ broadly reproduce the DNS mean velocity, both in amplitude and location of key features, with the best fit near the wall ($z = 1$), and the worst near the channel centerline ($z=0$).  Subsequently approximating $b_{ij,\mathrm{DNS}}$ by $b_{ij,\mathrm{TBRF}}$ causes additional errors, but theses errors are of similar magnitude to the errors already made in the propagation.  In particular, key features are correct, and amplitudes are appropriate.  What is also clear however, is that predictions are still far more accurate than both non-linear eddy-viscosity models.  The model from \cite{Shih1993} is able to predict the location of the peaks of the in-plane flow magnitude quite accurately, but significantly underpredicts the overall magnitude. Predictions by the cubic eddy-viscosity model from \cite{Lien1996} are far off overall. 
%
\subsubsection{Backward Facing Step}
\begin{figure}
\centering
    \includegraphics[trim={0cm 0.0cm 0cm 0cm},clip,width=1\textwidth]{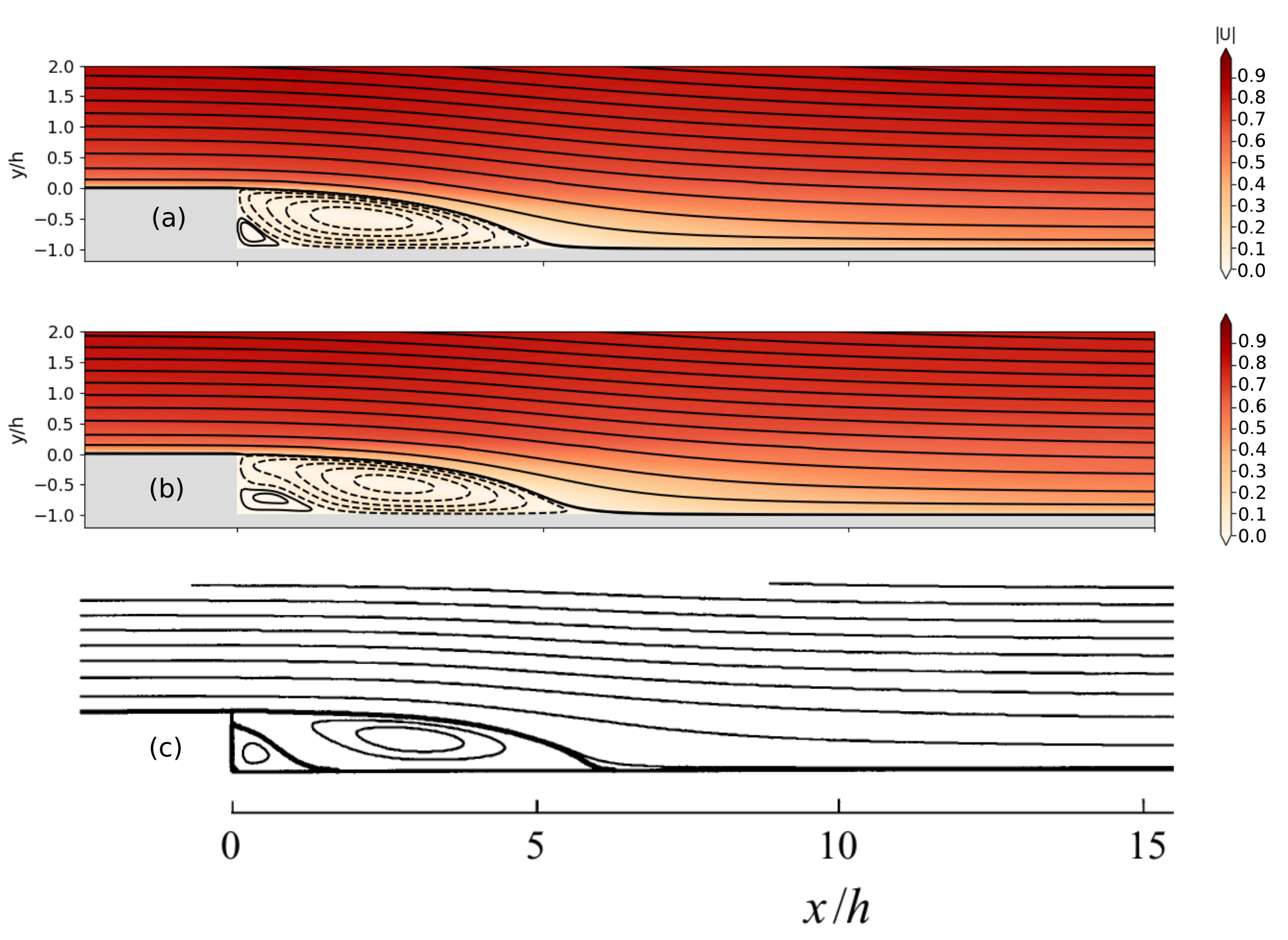}  
    \caption{Comparison of the streamlines for the backward facing step as given by (\textit{a}) the RANS $k-\omega$ simulation, (\textit{b}) the propagated flow field using $b_{ij,\mathrm{TBRF}}$, and (\textit{c}) DNS, adapted from \cite{Le1997}.}
    \label{fig:res_propBFS}
\end{figure}

Figure \ref{fig:res_propBFS} presents streamlines in the flow field for the backward facing step, with results from $k-\omega$ RANS, the propagated velocity field using the predicted anisotropy tensor from the TBRF algorithm ($b_{ij,\mathrm{TBRF}}$), and DNS data from \cite{Le1997}. The size of the recirculation region is more correctly predicted for the propagated velocity field using $b_{ij,\mathrm{TBRF}}$ compared to the baseline RANS simulation. Further away from the wall the solver using $b_{ij,\mathrm{TBRF}}$ does not introduce spurious effects and results are similar to the baseline RANS simulation.  The reattachment point locations ($x_\mathrm{reattach}$) for all three cases are presented in Table \ref{tab:res_BFS_xreatt}. A significant improvement is shown for the propagated velocity field compared to the baseline RANS simulation.
%

\begin{table}
  \begin{center}
\def~{\hphantom{0}}
\begin{tabular}{ll}
  \hline
    Model & $x_\mathrm{reattach}$ [x/h] \\\hline
    RANS & $5.45$\\
    RANS+$b_{ij,\mathrm{TBRF}}$ & $6.32$\\
    DNS \citep{Le1997}&  $6.28$ \\
    Experiment \citep{Jovic1994} & $6.0 \pm 0.15$\\
  \hline
  \end{tabular}
  \caption{Backward facing step, reattachment point locations.}
  \label{tab:res_BFS_xreatt}
  \end{center}
\end{table}

\begin{figure}
\centering
    \includegraphics[width=1\textwidth]{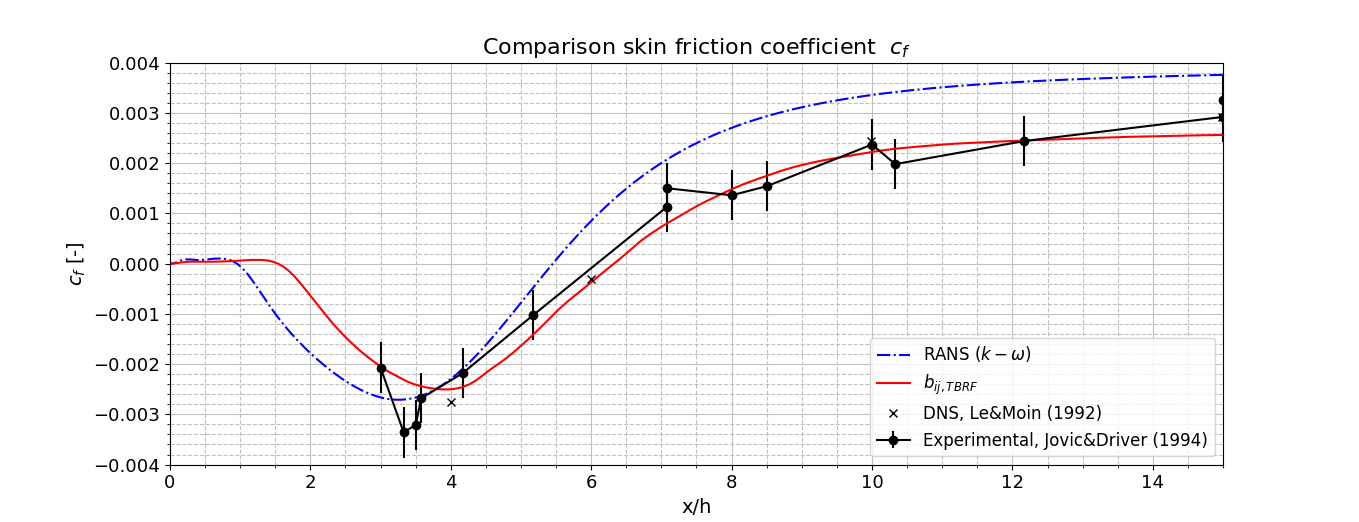}  
    \caption{Skin-friction coefficient for the backward facing step.  Experimental data from \cite{Jovic1994}; DNS data from \cite{Le1992}.}
    \label{fig:res_BFS_wallShearStress}
\end{figure}

%
%
The skin friction coefficients from the RANS simulation and the propagated flow field using $b_{ij,\mathrm{TBRF}}$ are compared to experimental data from \cite{Jovic1994} in Figure~\ref{fig:res_BFS_wallShearStress}. The propagated flow field shows a very close match to the experimental data, and the majority of results fall within the error bounds given by the experiment ($\pm 0.0005 c_f$). The reattachment point of the propagated flow field ($6.32$) compares favourably to the experimentally found reattachment point ($6.0 \pm 0.15$).
%
\section{Conclusions}
\label{sec:conclusions}
In this work, a novel random forest algorithm was introduced for RANS turbulence modeling, to model the Reynolds stress anisotropy tensor.  The algorithm was trained using invariant input features from several RANS ($k-\omega$) flow fields, and the corresponding responses for the anisotropy tensor from DNS or highly-resolved LES data. Galilean invariance of the predicted anisotropy tensor is ensured by making use of a tensor basis, derived in \cite{Pope1975}. The new random forest algorithm is called the Tensor-Basis Random Forest (TBRF) algorithm, similarly to the Tensor-Basis Neural Network from \cite{Ling2016} from which it was inspired. Robust predictions of the Reynolds-Stress anisotropy tensor are obtained by taking the median of the Tensor-Basis Decision Tree (TBDT) predictions inside the TBRF. 

Predictions for the Reynolds-stress anisotropy tensor were presented for the square duct flow case, curved backward-facing step, and backward-facing step.  Improvement is observed with respect to the baseline $k-\omega$ simulations, and the TBRF algorithm performs on par with the TBNN algorithm.  Compared to TBNN, the TBRF algorithm is relatively easy to implement and train (one does not have to think about matters such as the optimization algorithm used to tune the neural network weights and its convergence); the out-of-bag samples from the decision trees allow for a natural way to quantify the validation error during training and thus selecting the amount of trees to be used in the random forest. The few remaining hyperparameters are quite robust: the TBRF works well out-of-the-box even when using standard hyperparameter settings (fully grown trees, using all available features for creating the decision tree splits). 

A custom solver for propagating the anisotropy tensor was introduced, which blends the predictions for the anisotropy tensor with a $k-\omega$ turbulence model. This solver greatly increases numerical stability of the propagation.  Propagations for the square duct flow case and backward facing step are presented, which show a close match with respect to corresponding DNS and experimental data-sets. 

A possibility for future work might be using the TBRF for quantifying uncertainty of the predictions as well. For every location in the flow domain the trees in the random forest can be analyzed for their variance, which would make it possible to use an anisotropy eigenvalue perturbation methodology to quantify uncertainty such as proposed in \cite{Emory2013}. In order to achieve meaningful bounds for uncertainty of the predictions, one could look at e.g. Bayesian Additive Regression Trees \citep{Chipman2010}, or jackknife/infinitesimal jackknife methods \citep{Giordano2019}, or modify the random forest algorithm itself for meaningful uncertainty bounds, see e.g. \cite{Meinshausen2006, Mentch2016}. It was observed, that often the individual decision trees show a high variance of the prediction, when the prediction itself is relatively poor \citep{Kaandorp2018}.

Future work should further investigate the performance of the TBRF algorithm and the relaxation solver on more complex flow cases. Only a number of idealized canonical flow cases were considered here, it would be interesting to see how well the algorithm does in for example highly detached flows, how well the algorithm is able to extrapolate to higher Reynolds numbers, and whether it could be used for unsteady RANS flows as well. 

\section*{Acknowledgments}
The authors would like to thank Javier Fatou G\'{o}mez for supplying the OpenFOAM continuation solver used to propagate the Reynolds stresses into the flow field; also Julia Ling for providing her implementation of TBNN and support for this work.

\bibliographystyle{elsarticle-harv} 
\bibliography{library}

\newpage
\section*{Appendix: TBRF implementation details}
\label{app:1}

Decision trees base their predictions on a series of if-then tests on the input. Random forests consist of collections of decision trees with some randomized component differentiating trees.  Multiple decision tree algorithms exist, of which the CART (Classification And Regression Tree) algorithm is used as a starting point here.

As mentioned in Section \ref{subsec:TBRF}, the feature space is recursively split into two bins, $R_R$ and $R_L$.  The splitting is performed greedily, with each split selected to minimize the mismatch between the response $\yf$, and the best constant approximation of the response in both bins. Specifically for each split we solve \citep{Hastie2008}:
\begin{equation}
\small
\min_{j,s} \left[ \min_{c_L\in\mathbb{R}} \sum_{x_i \in R_L(j,s)} (\yf_i - c_L)^2 + \min_{c_R\in\mathbb{R}} \sum_{x_i \in R_R(j,s)} (\yf_i - c_R)^2 \right],
\label{eq:app_DecisionTree_eq}
\end{equation}
where $\yf_i$ denotes the response at $\Xf_i$.  Finding constants $c_L,c_R\in\mathbb{R}$ amounts to averaging $\yf$ within $R_L$ and $R_R$ respectively, effectively minimizing the variance in both bins.  Starting from the full data-set the same method is then applied to $R_L$ and $R_R$ in a recursive fashion.  The procedure is terminated either at a specified maximum branching depth, or a minimum number of samples per bin.

The new TBDT algorithm is comparable with the CART decision tree algorithm, but instead of approximating the response with constant values $c_L$ and $c_R$, the algorithm finds a constant value for each of the tensor basis coefficients $g^{(m)}$ in \eqref{eq:TensorBasis_Ling}, chosen to minimize the mismatch between this expression and the anisotropy tensor from DNS.  Specifically we solve
\begin{equation}
\min_{j,s} \left[ \min_{g_L^{(m)} \in \mathbb{R}^{10}} \sum_{x_i \in R_L(j,s)} \left\| \sum_{m=1}^{10} \Tf_i^{(m)} g_L^{(m)} - \mathsfbi{b}_i \right\|_F^2 + \min_{g_R^{(m)} \in \mathbb{R}^{10}} \sum_{x_i \in R_R(j,s)} \left\| \sum_{m=1}^{10} \Tf_i^{(m)} g_R^{(m)} - \mathsfbi{b}_i \right\|_F^2 \right],
\label{eq:app_TBDT_algor1}
\end{equation}
where $i$ indexes the samples, $\mathsfbi{b}_i$ is the DNS/LES anisotropy tensor, and $g_L^{(m)}$ and $g_R^{(m)}$ are the tensor basis coefficients in the left and right bins respectively.  The norm used is the Frobenius norm:
\begin{equation}
\|\Af\|_F := \sqrt{\sum_{i,j}  [\Af]_{ij} ^2} = \sqrt{\mathrm{tr}(\Af^T\Af)},
\label{eq:app_tensorNorm}
\end{equation}
where $\Af$ can be an arbitrary second-order tensor. This norm is chosen for its invariance to unitary transformations -- in particular rotations.
Now finding $g_L^{(m)}$ and $g_R^{(m)}$ amounts to solving two least-squares problems.
This must be repeated for every $j$ and for every optimization iteration of $s$.  As for
CART, \eqref{eq:app_TBDT_algor1} is repeated for each bin, until a stopping criterion is
reached.


Explicitly, by flattening the tensor at each point, and defining:
\begin{equation}
\hat{\Tf}_i = \left[ \begin{array}{cccc}
  [\Tf_i^{(1)}]_{11} & [\Tf_i^{(2)}]_{11} & \cdots & [\Tf_i^{(10)}]_{11} \\
  \lbrack\Tf_i^{(1)}\rbrack_{12} & [\Tf_i^{(2)}]_{12} & \cdots & [\Tf_i^{(10)}]_{12} \\
  \vdots  & \vdots  & \ddots & \vdots  \\
  \lbrack\Tf_i^{(1)}\rbrack_{33} & [\Tf_i^{(2)}]_{33} & \cdots & [\Tf_i^{(10)}]_{33} 
 \end{array} \right], \quad
\hat{\bfff}_i = \left[ \begin{array}{c}
 \lbrack \bfff_i\rbrack_{11} \\
 \lbrack \bfff_i\rbrack_{12}  \\
 \vdots    \\
 \lbrack \bfff_i\rbrack_{33}
\end{array}\right],
\label{eq:app_TBDT_algor4}
\end{equation}
each of the minimization problems over $\gf$ becomes $\min_\gf J$ where
\begin{equation}
J = \sum_{i=1}^N \| \hat{\Tf}_i \gf - \hat{\boldsymbol{b}}_i \|^2,
\label{eq:app_TBDT_algor3}
\end{equation}
with solution
\begin{equation}
\gf = \left(\sum_{i=1}^N \hat{\Tf}_i^T \hat{\Tf}_i \right)^{-1}\left(\sum_{i=1}^N \hat{\Tf}_i^T \hat{\boldsymbol{b}}_i\right).
\label{eq:app_TBDT_algor5}
\end{equation}
which can be solved separately for $R_L$ and $R_R$ to obtain $g_L^{(m)}$ and $g_R^{(m)}$.  The overall cost of this algorithm (as for CART) is domainated by sorting the data-values with respect to coordinate $j$.  This cost is $\mathcal{O}(N\log N)$ in the number of data-values, leading to an overall cost of $\mathcal{O}(N\log^2 N)$.  Unlike training neural networks, this procedure is fast, robust, easy to implement, and independent of any starting guess.

Due to the redundancy of the tensor-basis for any given sample $i\in\{1,\dots,N\}$, this problem can become ill-posed, especially towards the leaves of the tree, when only a few samples remain in a bin.  Therefore some $L^2$-regularization is added to $J$ with coefficient $\Gamma\in\mathbb{R}^+$, c.f.\ ridge regression.  To summarize, \eqref{eq:app_TBDT_algor5} is modified to
%
%
\begin{equation}
\gf = \left(\sum_{i=1}^N \hat{\Tf}_i^T \hat{\Tf}_i + \Gamma\If \right)^{-1}\left(\sum_{i=1}^N \hat{\Tf}_i^T \hat{\boldsymbol{b}}_i\right).
\label{eq:app_TBDT_algor6}
\end{equation}
In practice it was observed that by taking the median of all decision trees instead of the mean (see Section \ref{subsec:TBRF}), this already provides a lot of robustness, and that $\Gamma$ can be set to a very low value in general. The value of $\Gamma$ in the random forest was tuned using validation data-sets, see Section~\ref{subsec:res_predictions}.



%

%

\begin{figure}
\centering
    \includegraphics[trim={1.5cm 0cm 1.2cm 0cm},clip,width=1.0\textwidth]{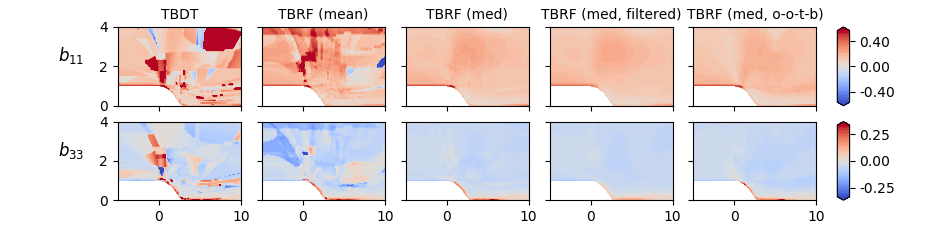}  
    \caption{Illustration of the robustness of the TBRF algorithm.}
    \label{fig:robustness}
\end{figure}

One important difference of the TBRF over the standard random forest is the fact that it effectively does not directly predict the final outcome, but the coefficients $\gf$, which multiply the basis tensors $\Tf$. Unlike the standard random forest, this means that the values for the final predictions do not have to lie in-between the values of the points used for training.  Decision trees are well known to be sensitive to small changes in the data, and this manifested during testing as highly irregular and inconsistent predictions in small regions of the spatial domain.  For a sample TBDT prediction of two components of the anisotropy tensor for the curved backward facing step (further introduced in Section~\ref{subsec:FlowCases}), where these inconsistencies are clearly visible, see the first column in Figure~\ref{fig:robustness}.  Both pruning and regularization were applied to reduce these inconsistencies. While regularization fixed the problem to some extent, it worsened predictions in certain regions of the flow, since this did not allow the coefficients $\gf$ to vary sufficiently.  Instead, it proved to be more successful to take the median of the trees in the random forest instead of the mean (as for example investigated in \cite{Roy2012}), since this removed sensitivity to outliers in the predictions. A comparison between taking the mean and median is presented in column 2 and 3 of Figure~\ref{fig:robustness}. Column 4 presents the results after applying the Gaussian smoothing as described in Section~\ref{subsec:TBRF}. As described in Section~\ref{subsec:TBRF}, the TBRF algorithm will work out-of-the-box without much tuning quite well, as demonstrated in column 5 of Figure~\ref{fig:robustness}. Predictions are shown for a TBRF with standard settings (fully grown decision trees, using all features for creating the splits), and a arbitrarily low regularization factor of $\Gamma = 1 \times 10^{-15}$. As can be seen the anisotropy predictions are quite insensitive to the set of hyperparameters when comparing to the predictions from the tuned TBRF in column 3.

\end{document}